\pdfoutput=1   
\documentclass[usenatbib,usegraphicx,fleqn]{ctextemp_mn2e}
\usepackage{amsmath,amssymb,bm,color,multirow,multicol}
\usepackage{charter}   

\newcommand{\e}[1]{\times 10^{#1}}

\usepackage{ulem}

\def\dst{\displaystyle}

\title[Warps, Bending and Density Waves Excited by Rotating Magnetized Stars:]{Warps, Bending and Density Waves Excited by Rotating Magnetized Stars:
Results of Global 3D MHD Simulations}

\author[M. M. Romanova et al.]
{M. M. Romanova,$^1$\thanks{e-mail:romanova@astro.cornell.edu}, G. V. Ustyugova$^2$,
A. V. Koldoba$^2$, R. V. E. Lovelace $^{1}$\\
$^1$ Department of Astronomy, Cornell University, Ithaca, NY 14853-6801, USA\\
$^2$ Keldysh Institute for Applied Mathematics, Moscow, Russia\\}

\begin{document}

\maketitle

\begin{abstract}

\noindent  We report results of the first global three-dimensional (3D)
magnetohydrodynamic (MHD) simulations of the waves excited in an accretion disc by a
rotating star with a dipole magnetic field misaligned from the star's rotation axis
(which is aligned with the disc axis). The main results are the following: (1) If the
magnetosphere of the star corotates approximately with the inner disc, then we observe
a strong one-armed bending wave (\textit{a warp}).
   This warp  corotates with the star and has a maximum amplitude between corotation radius and
   the radius of the vertical resonance. The disc's center of mass can deviate from
   the equatorial plane up to the distance of $z_w\approx 0.1 r$. However, the effective height of
   the warp can be larger, $h_w \approx 0.3 r$
   due to the finite thickness of the disc.
Stars with a range of misalignment angles excite warps.
    However, the amplitude of the warps is larger for misalignment angles between $15$ and $60$ degrees.
    The location and amplitude of
the warp does not depend on viscosity, at least for relatively small values of the
standard alpha-parameter, up to $0.08$. (2) If the magnetosphere rotates slower, than
the inner disc, then a bending wave is excited at the disc-magnetosphere boundary, but
does not form a large-scale warp.
    Instead, persistent, high-frequency oscillations become strong at the inner
region of the disc.
   These are (a) \textit{trapped density waves} which form
inside the radius where the disc angular velocity has a maximum,  and  (b)
\textit{inner bending waves} which appear in the case of accretion through magnetic
Rayleigh-Taylor instability. These two types of waves are connected with the inner
disc and their frequencies will vary with accretion rate. Bending oscillations at
lower frequencies are also excited including global oscillations of the disc. In cases
where the simulation region is small, slowly-precessing warp forms with the maximum
amplitude at the vertical resonance. The present simulations are applicable to young
stars, cataclysmic variables, and accreting millisecond pulsars.  A large-amplitude
warp of an accretion disc can periodically obscure the light from the star. Different
types of waves  can be responsible for both the high and low-frequency quasi-periodic
oscillations (QPOs) observed in different types of stars. Inner disc waves can also
leave an imprint on frequencies observed in moving hot spots on the surface of the
star.

\end{abstract}


\section{Introduction}

Different types of accreting stars have dynamically important magnetic fields, such as
young T Tauri stars (e.g., \citealt{bouvier07a}), accreting millisecond pulsars (e.g.,
\citealt{vanderKlis00,vanderKlis06}), and accreting white dwarfs (e.g.,
\citealt{warner95,warner04}).


If a rotating star has a tilted dipole (or other non-axisymmetric) magnetic field then
it applies a periodic force on  the inner part of the accretion disc and different
types of waves are excited and propagate to larger distances (e.g., \citealt{lai99}).
     The linear theory of waves in discs has been developing
extensively over many years (see, e.g.,  book by \citealt{kato98} and references
therein).
     If an external force is applied to the disc  from the tilted dipole of
 a rotating star (or from a
 secondary star or planet), then this force can generate
 density waves in the disc (e.g., \citealt{gold78}) and
vertical bending waves (e.g., \citealt{lubow81,lai99,ogil99}).
     Theoretical investigations have led
to an understanding that both types of waves may be strongly enhanced or damped at
radii corresponding to resonances  (e.g., \citealt{kato98}).
      There are two types of  resonances:
horizontal resonance (Lindblad resonance) and  vertical resonance. Another important
resonance is associated with corotation radius $r_{\rm cr}$ where the angular velocity
of the disc matches the angular velocity of the external force.

\citet{aly80} and \citet{lipunov80} investigated the tilting of the inner disc due to
the magnetic force applied by the rotating  dipole magnetosphere of the star
where both the magnetic and rotational axes of the star are misaligned relative to the
rotational axis of the outer disc. They concluded that the magnetic force acting on
the inner disc causes it to strongly depart from the original plane of the disc.
\citet{lai99} investigated this case and concluded that the magnetic force also drives
a warp at the inner disc which slowly precesses around the rotational axis of the
disc.

\begin{figure}
\centering
\includegraphics[width=8.5cm]{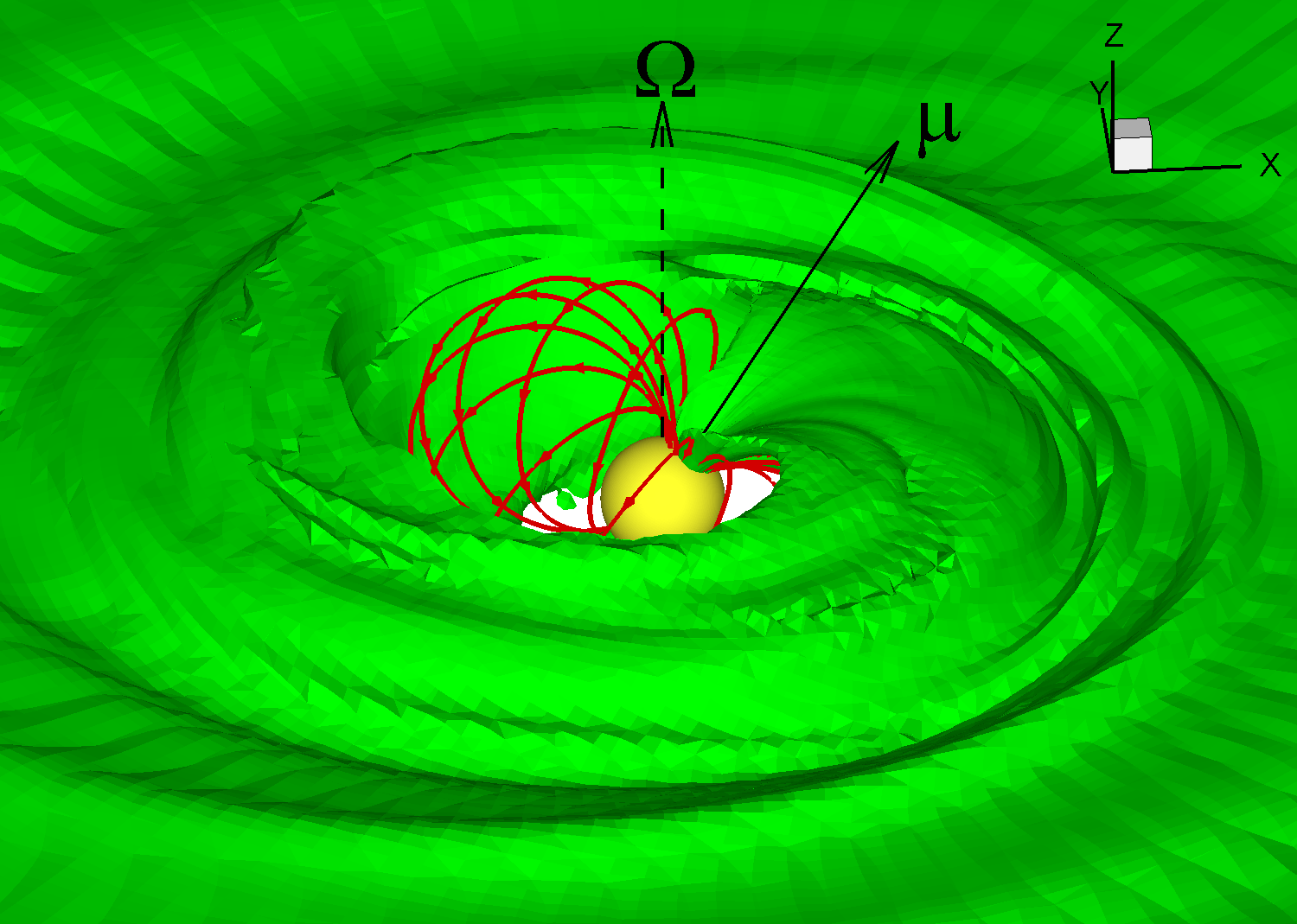}
\caption{An example of the magnetospheric accretion and a warp driven by the rotating
and tilted at $\theta=30^\circ$ dipole magnetosphere with $\tilde{\mu}=1.5$ (model
FW$\mu$1.5). The background shows one density level ($\rho=0.07$). The lines are
sample magnetic field lines of the closed magnetosphere. The vectors $\bf{\mu}$ and
$\bf{\Omega}$ show directions of the magnetic and rotational moments of the star.}
\label{warp-1}
\end{figure}

\citet{terquem00} investigated the formation of a disc warp in the case where the
rotational axis of the star is aligned with that of the disc, and the inner disc
corotates with the star, taking into account an $\alpha-$type viscosity in the disc
\citep{shakura73}.
   They concluded that for a
relatively small viscosity, $\alpha=0.01-0.08$, a warp is excited near the
magnetosphere and corotates with the star. However, it is suppressed by the viscosity
at larger distances from the star.
    They also concluded that the height
of the warp can reach up to $\sim 10\%$ of the radial distance. One of the main
motivations for studying inner disc warps comes from observations of young T Tauri
stars (CTTS) in which the light curves show regular dips that can be interpreted as
obscuration events by a warped disc (e.g., \citealt{bouvier99,bouvier03,bouvier07b},
also \citealt{carpenter01}).

Another important impetus to study waves in the disc comes from observations of
quasi-periodic oscillations (QPOs) in different accreting stars: neutron stars, black
holes, and white dwarfs.
 For example, in low-mass X-ray binaries (LMXBs), typical QPO
frequencies vary from very high, $\nu\approx 1300$Hz, down to very low, $\nu\sim
0.1$Hz (e.g., \citealt{vanderKlis06}).
    An important feature of the spectra is the presence of `twin peaks'
of high-frequency QPOs. In many cases, the separation between peaks is equal to either
the frequency of the star or half of its value (e.g., \citealt{vanderKlis00}).
   Different theories have been proposed to explain this phenomenon (e.g.,
\citealt{lamb85,mill98,love07}). In a number of accreting millisecond pulsars, the
difference between peaks does not correlate with the frequency of the star and varies
in time significantly (e.g., Sco X-1), and this also requires an explanation (e.g.,
van der Klis et al. 1997).

In accreting magnetized stars, the magnetosphere may rotate slower than the inner disc
(e.g., \citealt{roma02}, \citealt{long05}).
    Consequently, there is a  maximum in
the angular velocity distribution as a function of radius at the inner disc. This
condition can lead to formation of unstable radially trapped Rossby waves
\citep{love07,love09}. These waves can be responsible for one of the high-frequency
QPO peaks observed in LMXBs. These waves may play a role of blobs suggested earlier in
theoretical
 models of QPOs, such as in the beat-frequency model (e.g., \citealt{mill98,lamb03}). However, compared with the
blobs, the inner density waves are much more ordered, and have higher coherence.
However, the origin of the second QPO peak is still unclear.

Different QPO frequencies  in discs around stars and black holes can be connected with
 resonances in the disc, where the amplitude of waves may be enhanced (e.g.,
\citealt{kato98,kato04,zhang06}).
     For example,
\citet{lamb03} and \citet{lai08} suggest that wave can be excited at the spin-orbital
resonance, and the light from the high-frequency QPO can interact with this wave and
result in the lower-frequency QPO.
     \citet{kato04} suggested that in the case of the tilted black holes the
disc may be warped, and investigated resonance associated with non-linear interaction
of different waves with this warp (see also \citealt{petri05,lai08,kato08,kato10}).
Slowly-precessing warps can form in both the black hole and the magnetic star cases
and can be responsible for low-frequency QPOs (e.g., \citealt{lai99,pfeiffer04}).

The very low-frequency oscillations ($\sim 0.1-1$Hz) in LMXBs can be connected with
disc oscillations at large distances form the star (e.g., corrugation waves can be
excited, \citealt{kato98}).
Such global oscillations have been observed in numerical
simulations of accretion to a black hole with a tilted spin axis
\citep{fragile05,fragile07}.
    The authors note that the frequency of
these oscillations depends strongly on the size of the disc.

In all above-mentioned cases, where a rotating magnetized star excites waves in a
disc, the waves have been analyzed using linearized equations.
     The linear theory provides an important
theoretical understanding of the different waves.
    However, the approach has significant limitations,
particularly in those cases where the amplitude of waves is not small.
     Further, the theory gives only a rough treatment  of the
force on a disc due to a tilted stellar magnetic field. In this paper, we present the
first results of global three-dimensional (3D) MHD simulations of waves excited by a
rotating star with a misaligned dipole magnetic field. Earlier, we investigated the
magnetospheric flow around tilted dipoles (e.g., \citealt{roma03,roma04}), and motion
of hot spots at the stellar surface (e.g., \citealt{roma08,kulk09,bac10}). However,
this is the first time when we investigate waves in the disc which are excited by a
rotating tilted dipole.

In Sec. \ref{sec:model-all} we briefly describe the linear theory of waves in discs,
our numerical model and methods for wave analysis in simulations. We also discuss the
theory of waves in Keplerian discs in Appendix \ref{sec:Appendix-allwaves-modes} and
the theory of trapped waves in Appendix \ref{sec:Appendix-trapped modes}. In Sec.
\ref{sec:results-main} we present the results of our simulations. We discuss
applications of the numerical results in Sec. \ref{sec:applications-discussion}.
Conclusions are given in Sec. \ref{sec:conclusions}.

\section{The model, and analysis of waves in the disc}
\label{sec:model-all}

Below we
give the highlights of the linear theory of waves (Sec. \ref{sec:theory}), describe
our numerical model (Sec. \ref{sec:nummodel-refunits}) and the approach which has been
used for analysis of waves in numerical simulations (Sec. \ref{sec:theory-analysis of
waves}).

\subsection{Theoretical background} \label{sec:theory}

   Here we briefly summarize the theory of waves excited in accretion discs
(see, e.g., \citealt{kato98}). In the majority of theoretical studies, the disc is
approximated to be thin, isothermal and non-magnetic. Furthermore, oscillations can be
free, or can be excited by the external force. Below we briefly discuss different
types of waves.

\subsubsection{Free oscillations in the disc}

 Small perturbations of a thin, non-magnetized disc
involve in-plane and out-of-plane displacements of the disc matter. Waves can be
roughly divided into inertial-acoustic waves (referred to as p-modes, where the
restoring forces are rotation and  pressure gradient), and and inertial-gravity waves,
referred to as g-modes where the restoring force is gravity\footnote{Note that the
perturbations are also assumed to be isothermal, and hence the buoyancy forces are
neglected ($N^2 = 0$).} Inertial-acoustic modes are in-plane modes, and they propagate
in the form of pressure or density disturbances. They are often termed ``density
waves" (see also \citealt{gold78}).
 The inertial-gravity modes are
out-of-plane modes \citep{kato98}. \textbf{These waves lead to misplacement of the
local center of mass of the disc  out of the equatorial plane and hence are called
``bending waves".}

For both the in-plane and out-of-plane modes, the azimuthal dependence is
$\exp(im\phi)$, with $m=0,~1,~2,..$. The $m=1$ in-plane perturbation is a one-armed
spiral wave and the $m=2$ in-plane mode is a two-armed spiral wave.
    The out-of-plane modes involve
a small vertical displacement or shift of the midplane of the disc, $z_w(t,r,\phi)$
(cylindrical coordinates) with $|z_w| \ll r$.

To investigate the propagation of small-amplitude waves in the  disc,
 perturbations of the hydrodynamical values  $Q = p,~
\rho, ~v_r, ..$ are expanded as (e.g., \citealt{okazaki87,tanaka02,lai08}):
\begin{equation}
\dst Q(r,z,\phi,t) = \sum_{m,n,\omega} Q_{m,n,\omega} (r) H_n\left(\frac{z}{h}\right)
\exp(im \phi - i\omega t)~.
\end{equation}
Here,  $\omega$  is the angular frequency of the wave, $m=0,~1,~2...$  is the number
of arms in the azimuthal direction, $H_n, n \ge 0$  are the Hermite polynomials (with
$n=0,~1,~2,..$), and $h(r)$ is the half-thickness of the disc at radius $r$. Waves
with $n=0$ and $n=1$ represent density and bending waves, respectively.

Linearization of the equations of motion (e.g., \citealt{kato98}) leads to
\begin{equation}
 \frac{d^2 \psi}{dr^2} + \frac{(\tilde\omega^2 - \kappa^2)
(\tilde\omega^2 - n\Omega_\perp^2)}{c_s^2 \tilde \omega^2} \psi = 0~,
\label{eq:main-wave}
\end{equation}
where  $\psi = \delta p/\rho_0$ is the perturbation of the enthalpy for density waves
or $z_w$ for the bending waves.
  Also in this equation, $\tilde\omega
\equiv \omega -m \Omega$ is the Doppler-shifted angular frequency of the wave seen by
an observer orbiting at the disc's angular velocity $\Omega(r)$, $\Omega_\perp$ is the
frequency of oscillations perpendicular to the disc, $\kappa^2=r^{-3}
d(r^4\Omega^2)/dr$ is the square of the radial epicyclic frequency, and $c_s$ is the
sound speed in the disc.

The local free wave solution in the
 WKBJ approximation can be considered as (\citealt{kato98}),
\begin{equation}
\psi \propto k_r^{-1/2}\exp\big[i\int^r dr k_r(r)\big]~,
\end{equation}
for  $(k_r r)^2\gg1$ and $(k_rh)^2\ll1$, where $k_r$ is the radial wavenumber.
    Thus, equation (\ref{eq:main-wave}) can be presented as  $d^2\psi/dr^2 = -k_r^2 \psi$
 and one gets the
 dispersion relation:
\begin{equation}
 \frac{(\tilde\omega^2 - \kappa^2)
(\tilde\omega^2 - n\Omega_\perp^2)}{\tilde \omega^2} = k_r^2 c_s^2~.
\label{eq:main-wave-disperrelation}
\end{equation}
Propagation of the perturbations is possible in regions  where $ (\tilde\omega^2 -
\kappa^2) (\tilde\omega^2 - n \Omega_\perp^2) > 0$.
    The WKBJ approximation breaks down in the vicinity of points
where the coefficient of $\psi$ in equation (\ref{eq:main-wave}) changes sign or has a
singularity.
      At the points where $\tilde\omega = \pm \kappa$ there is an outer (+sign)
      or an inner (-sign) Lindblad resonance.   At the points where $\tilde\omega= \pm
n\Omega_\perp$ (for $n \ne 0$) there are vertical resonances. For $\tilde\omega= 0$
there is a corotation resonance.
   Appendix \ref{sec:Appendix-allwaves-modes} summarizes the behavior of these
different modes.

     In this work we are mainly interested  in (1) oscillations in the
plane of the disc ($n=0$), and (2) out of plane oscillations  ($n=1$).
    Further, we consider only the
one-armed ($m=1$) and two-armed ($m=2$) modes.
   If the accretion disc is Keplerian, then  $\kappa=\Omega_\perp=\Omega$.
  From the analysis of \citet{zhang06}, one-armed
density waves ($m=1$) at $n=0$ can propagate only in the region of the disc outside
the outer Lindblad resonance.
   The inner Lindblad and corotational
resonances in this case ($n=0$) are absent.
   For vertical oscillations ($n=1$), the
position of the outer Lindblad and vertical resonances coincide.
        Two-armed density waves ($m=2$) with $n=0$ can
propagate either in the external parts of the disc outside the outer Lindblad
resonance or in the region inside the inner Lindblad resonance.

\begin{table}
\begin{tabular}{llll}
\hline                & CTTSs              & White dwarfs   & Neutron stars      \\
\hline
$M(M_\odot)$          & 0.8                & 1              & 1.4            \\
$R_*$                   & $2R_\odot$         & 5000 km        & 10 km              \\
$B_*$ (G)             & $10^3$             & $10^6$         & $3\e8$             \\
$R_0$ (cm)            & $4\e{11}$          & $1.4\e9$       & $2.9\e6$           \\
$v_0$ (cm s$^{-1}$)   & $1.6\e7$           & $3\e8$         & $8.1\e9$           \\
$\rho_0$ (g cm$^{-3}$) & $2.8\e{-11}$       & $7.9\e{-8}$      & $1.0\e{-5}$           \\
$\sigma_0$ (g cm$^{-2}$) & $11.0$       & $112.5$          & $28.9$           \\
$\nu_0$               & $0.55$ day$^{-1}$  & $3.2\e{-2}$ Hz & $4.5\e2$ Hz        \\
$P_0$                 & $1.8$ days         & 29 s           & 2.2 ms             \\
\hline \label{tab:refval}
\end{tabular}
\caption{Sample values of the physical parameters of different types of stars. See
Sec. \ref{sec:nummodel-refunits} for a description.} \label{tab:refval}
\end{table}

\subsubsection{Waves excited by the external force. Resonances}
\label{sec:theory external force}

In our model, waves are `driven'  by the rotating tilted magnetosphere of the star.
The interaction of the magnetosphere with the disc is a complex, non-linear, and
non-stationary process. For simplicity of the analysis, this interaction can be
approximated by the force acting on the disc and presented as a sum of different
harmonics:
\begin{equation}
f = f (r,z,\phi - \Omega_* t) = \sum_m f_m(r,z) \exp[im(\phi - \Omega_*t)]~,
\label{eq:force}
\end{equation}
where $\Omega_*$ is the angular frequency of the star. In linear approximation, we
suggest that the $m-$harmonic of this force excites the $m-$armed wave in the disc.
The frequency of this force is $m\Omega_*$. Then, for a Keplerian disc, the condition
for the Lindblad resonances $\tilde\omega=\pm\Omega$ will be
$m\Omega_*-m\Omega=\pm\Omega$, or
\begin{equation}
r_{LR}=r_{cr}\bigg(1\pm\frac{1}{m}\bigg)^{2/3}~,
\end{equation}
where $r_{cr}= (GM/\Omega_*^2)^{1/3}$ is the corotation radius. For bending waves,
$n=1$, the condition for  vertical resonances $\tilde\omega=\pm\Omega$ is the same,
and hence the vertical resonances are located at the same radii.

    For a one-armed density
wave ($m=1$, $n=0$) in a Keplerian disc, the outer Lindblad resonance is located at
radius $r_{\rm OLR}= (4GM/\Omega_*^2)^{1/3}$.
   The location of the outer vertical resonance
$r_{\rm OVR}$ for a bending wave
 ($n=1$) is the same.
For a two-armed spiral wave ($m=2$), the outer Lindblad and vertical resonances are
located at the distance of $r_{\rm OLR}=r_{\rm OVR}=[9GM/(4 \Omega_*^2)]^{1/3}$, while
the inner Lindblad resonance is at the distance  $r_{\rm ILR}=r_{\rm IVR} =
[GM/(4\Omega_*^2)]^{1/3}$.
     The corotation resonance for all waves is located at radius
$r_{\rm CR}=(GM/\Omega_*^2)^{1/3}$. A more detailed description of the modes is given
in Appendix \ref{sec:Appendix-allwaves-modes}. In this paper we compare position of
waves observed in simulations with position of resonances derived theoretically.

\subsubsection{High-frequency trapped waves near magnetized stars}

In relativistic discs, the epicyclic frequency $\kappa$  has a maximum at the inner
disc (where relativity effects from a black hole or a neutron star are stronger), and
this may lead to formation of waves which are trapped below this maximum (e.g.,
\citealt{okazaki87,nowak91,kluzniak02}). These waves are important because they may be
responsible for the high-frequency QPOs observed in black hole-hosting systems (e.g.,
\citealt{stella99,alpar05,alpar08}).

In cases where a star has a dynamically-important magnetic field and is slowly
rotating, the magnetosphere slows down the rotation of matter at the inner part of the
disc. Hence, there is a maximum in the disc angular velocity distribution. This
situation is similar to that of relativistic discs: high-frequency waves can be
trapped below the maximum of the angular velocity distribution (e.g.,
\citealt{love07,love09}).
    We briefly summarize the theory
of trapped waves around magnetized stars in Appendix \ref{sec:Appendix-trapped modes}.

\subsubsection{Low-frequency bending waves}

In the absence of an external force, perturbation of the disc in the vertical
direction leads to the formation of free bending waves. Namely, the matter lifted
above the equatorial plane ($z=0$) to the height $z$ experiences the restoring gravity
force $g=-GM_*z/r^3$ per unit mass. That is why the frequency of matter oscillation in
the vertical direction is $\Omega_\perp\approx \Omega=\sqrt{GM/r^3}$, therefore, at
large distances from the star, these oscillations have a very low frequency. A one-arm
global bending wave with low frequency and long wavelength in the radial direction
(or, corrugation wave)
 can propagate to very large radial distances
\citep{kato98}. In addition, global bending oscillations of the \textit{whole disc}
are possible. Waves can also be reflected or generated at the outer boundary of the
disc \citep{zhang06}. Such low-frequency oscillations can be excited by the rotating
magnetosphere of the star, in particular if the star rotates slowly.

\subsubsection{More realistic discs}

Real accretion discs have a finite thickness, and hence the thin disc approximation
may not be applicable. In addition, the disc may have a complex distribution of
density, temperature and magnetic field. All these factors may determine the formation
and propagation of waves
(e.g., \citealt{kato98}, \citealt{fu-lai12}).

Waves in discs of finite thickness were investigated in few global 3D simulations of
accretion onto rotating black holes with the tilted spin. Formation of warp and tilted
discs have been observed in these simulations (e.g., \citealt{nelson00,fragile05}).
Different disco-seismic modes (such as trapped g-mode), were observed in axisymmetric
simulations of accretion on to a non-rotating black hole from non-magnetized accretion
disc (e.g., \citealt{o'neil09}). However, in discs where accretion is driven by the
magneto-rotational instability (MRI, e.g., \citealt{balbus91}), no interesting waves,
such as trapped relativistic g-modes were observed (e.g., \citealt{reynolds09}). These
simulations show that the magnetic field may possibly damp waves in the disc.

In our simulations, the disc has a finite thickness, $h/r\approx 0.1$, which
approximates a thin disc. We observed in simulations that the angular velocity is
almost Keplerian in major part of the disc (excluding the inner parts near the
magnetosphere) and hence the approximation for Keplerian disc is valid. Some magnetic
flux of the magnetosphere is trapped inside the disc and hence may influence the
positions of resonances. However, this is the region where the angular velocity is
strongly non-Keplerian, and where we can not apply the standard theory of waves.
Overall, results obtained in our model can be compared with results obtained with the
linear theory of waves in a thin disc, excluding the innermost region of the disc,
where the theory of trapped waves \citep{love07,love09} is more appropriate.

\begin{table*}
\centering
\begin{tabular}{llllllll}

\\ Name & $\widetilde{\mu}$ & $r_{\rm cor}$ & $\Omega_*$ & $\theta$ & $\alpha$ & $R_{\rm out}/R_*$ & Warp Properties\\
\hline
FW$\mu$0.5  & 0.5 & 1.5 & 0.54 & $30^\circ$ & $\alpha=0.02$ & 34.5 & fast, $\Omega_w\approx \Omega_*$\\
FW$\mu$1.5  & 1.5 & 1.8 & 0.41 & $30^\circ$ & $\alpha=0.02$ & 34.5 & fast, $\Omega_w\approx \Omega_*$\\
\hline
FW$\theta$5   & 0.5 & 1.5 &  0.54 & $5^\circ$  & $\alpha=0.02$ & 34.5 & fast, $\Omega_w\approx \Omega_*$ \\
FW$\theta$15  & 0.5 & 1.5 &  0.54 & $15^\circ$ & $\alpha=0.02$ & 34.5 & fast, $\Omega_w\approx \Omega_*$\\
FW$\theta$45  & 0.5 & 1.5 &  0.54 & $45^\circ$ & $\alpha=0.02$ & 34.5 & fast, $\Omega_w\approx \Omega_*$\\
FW$\theta$60  & 0.5 & 1.5 &  0.54 & $60^\circ$ & $\alpha=0.02$ & 34.5 & fast, $\Omega_w\approx \Omega_*$\\
FW$\theta$90  & 0.5 & 1.5 &  0.54 & $90^\circ$ & $\alpha=0.02$ & 34.5 & fast, $\Omega_w\approx \Omega_*$\\
\hline
FW$\alpha$0.00  & 0.5 & 1.5 & 0.54 & $30^\circ$  & $\alpha=0.00$  & 34.5 & fast, $\Omega_w\approx \Omega_*$ \\
FW$\alpha$0.04  & 0.5 & 1.5 & 0.54 & $30^\circ$ & $\alpha=0.04$ & 34.5 & fast, $\Omega_w\approx \Omega_*$\\
FW$\alpha$0.06  & 0.5 & 1.5 & 0.54 & $30^\circ$ & $\alpha=0.06$ & 34.5 & fast, $\Omega_w\approx \Omega_*$\\
FW$\alpha$0.08  & 0.5 & 1.5 & 0.54 & $30^\circ$ & $\alpha=0.08$ & 34.5 & fast, $ \Omega_w\approx \Omega_*$\\
\hline \hline
SWcor1.8  & 0.5 & 1.8 & 0.41 & $30^\circ$ & $\alpha=0.02$ & 34.5 & slow, $\Omega_w\approx 0.5 \Omega_*$  \\
SWcor3    & 0.5 & 3.0 & 0.19 & $30^\circ$ & $\alpha=0.02$ & 34.5 & slow, $\Omega_w\approx 0.1 \Omega_*$  \\
SWcor5    & 0.5 & 5.0 & 0.09 & $30^\circ$ & $\alpha=0.02$ & 34.5 & slow,  $\Omega_w\approx 0.02 \Omega_*$ \\
\hline
LRcor1.8  & 0.5 & 1.8 & 0.41 & $30^\circ$ & $\alpha=0.02$ & 57.7 & no big warp\\
LRcor3  & 0.5 & 3.0 & 0.19 & $30^\circ$ & $\alpha=0.02$ & 57.7 & no big warp\\
LRcor5  & 0.5 & 5.0 & 0.09 & $30^\circ$ & $\alpha=0.02$ & 57.7 & no big warp \\
\hline
SRcor3  & 0.5 & 3.0 & 0.19 &$30^\circ$ & $\alpha=0.02$ & 26.9 & disc oscillations, example\\
\hline
\end{tabular}
\caption{The main set of simulation runs. See  Sec. \ref{sec:simulation runs} for a
detailed description.} \label{tab:models}
\end{table*}

\begin{table*}
\centering
\begin{tabular}{llllllllllll}
\\  &  & & & $\bf{m=1}$ &  & &  & & $\bf{m=2}$ &&  \\
 \hline
   $\Omega_*$ & $r_{\rm CR}$ & & &  $r_{\rm OLR} $ & $r_{\rm OVR} $   & && $r_{\rm ILR}$ & $r_{\rm OLR}$ & $r_{\rm IVR}$ &  $r_{\rm OVR}$  \\
\\  &   &  &  & $n=0$&$n=1$ & & & $n=0$  &$n=0$  & $n=1$ &$n=1$ \\
\hline
$0.544$ & 1.5 & && 2.38 & 2.38 & && 0.94 & 1.97 & 0.94 & 1.97 \\
$0.414$ & 1.8 && & 2.86 & 2.86 && & 1.13 & 2.36 & 1.13 & 2.36 \\
$0.192$ & 3.0 && & 4.76 & 4.76 & && 1.89 & 3.93 & 1.89 & 3.93 \\
$0.089$ & 5.0 & && 7.93 & 7.93 && & 3.15 & 6.55 & 3.15 & 6.55 \\
\hline
\end{tabular}
\caption{Radii of  resonances in a thin hydrodynamic Keplerian disc excited by a
rotating perturbation from a magnetized star rotating with angular velocity $\Omega_*$
(in dimensionless units discussed in the text).  $r_{\rm CR}$ is the radius of the
corotation resonance.
   The next two columns show the radii of the outer
Lindblad  $r_{\rm OLR}$ and outer Vertical   $r_{\rm OVR}$  resonances in the case of
one-armed ($m=1$) density ($n=0$) and bending ($n=1$) waves.
    The right four columns show the radii for
resonances in the cases of two-arm  waves: inner and outer Lindblad resonances $r_{\rm
ILR}$ and $r_{\rm OLR}$ for density waves ($n=0$), and inner and outer vertical
resonances $r_{\rm IVR}$ and $r_{\rm OVR}$  for bending waves ($n=1$).}
\label{tab:resonances}
\end{table*}

\begin{figure*}
\centering
\includegraphics[width=16cm]{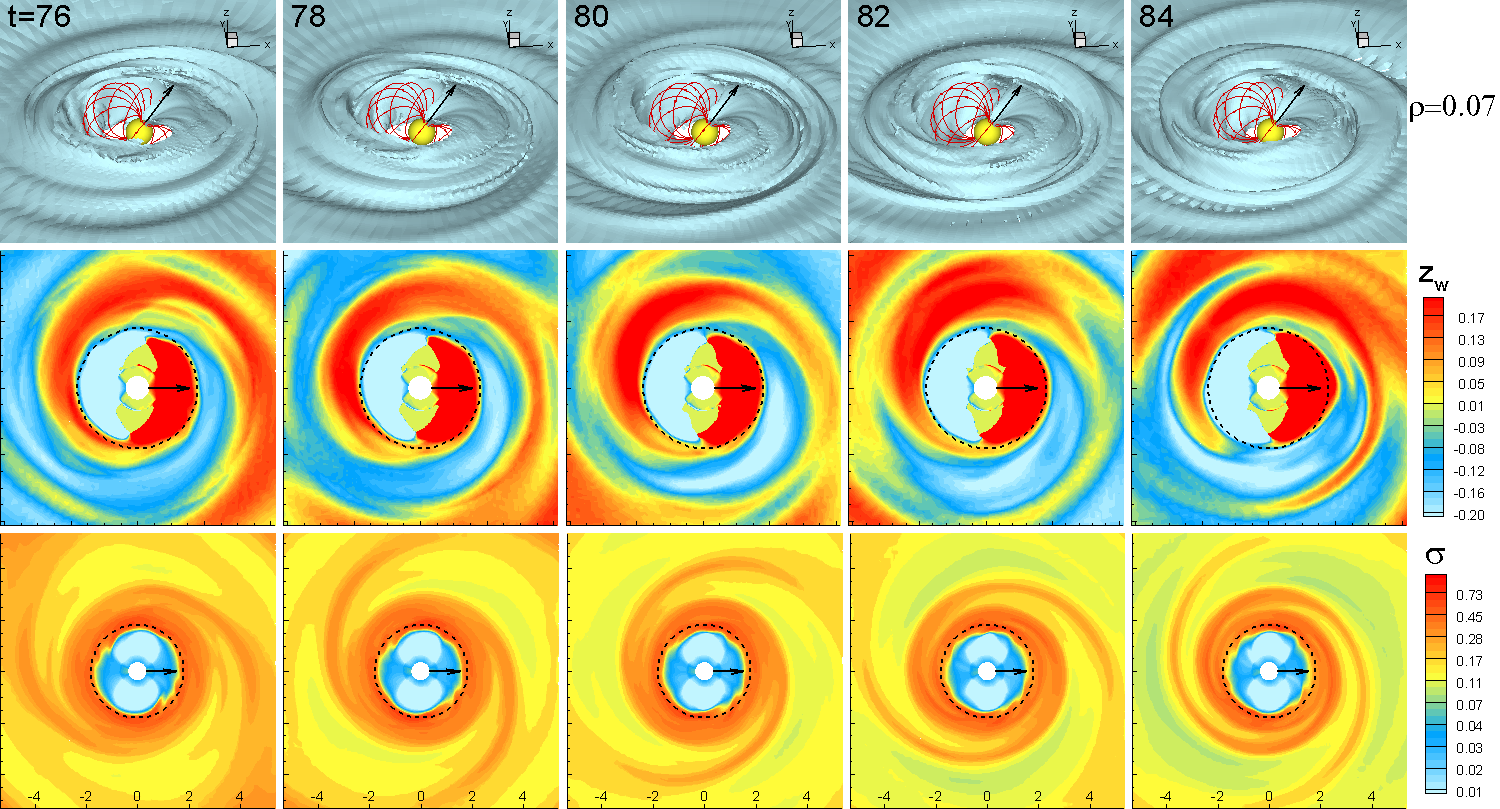}
\caption{Views of the inner parts of the disc in a model  with a relatively large
magnetic moment, $\widetilde{\mu}=1.5$ (model FW$\mu$1.5) at different times, $t$.
\textit{Top row:} A three-dimensional view of the disc at one density level,
$\rho=0.07$.   \textit{Middle row:} $z-$averaged height of the disc's center of mass
above or below the equatorial plane, $z_w$.   \textit{Bottom row:} surface density
distribution in the equatorial plane, $\sigma$. Time $t$ is given in units of $P_0/4$,
where $P_0$ is the orbital period at $r=1$. Plots are shown in a coordinate system
rotating with the star.} \label{warp-d1_5-c1_8-15}
\end{figure*}

\begin{figure*}
\centering
\includegraphics[width=16cm]{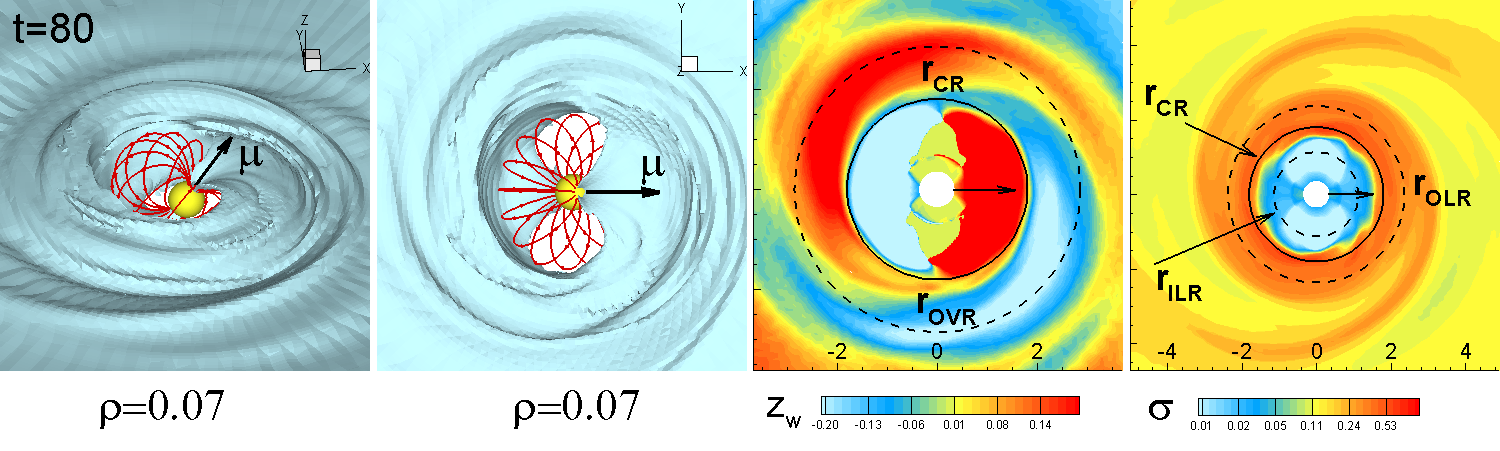}
\caption{Views of bending and density waves shown in Fig. \ref{warp-d1_5-c1_8-15} at
time $t=80$. The left two panels show  3D views of the warp from the side and face-on
orientations. The right two panels show bending and density waves and the locations of
resonances: $r_{\rm CR}$ is the corotation resonance;  $r_{\rm OVR}$ is the outer
vertical resonance (calculated for $m = 1$); $r_{\rm ILR}$ and $ r_{\rm OLR}$ are the
inner and outer Lindblad resonances respectively (calculated for $m = 2$). Plots are
shown in a coordinate system rotating with the star. Positions of resonances are taken
from Tab. \ref{tab:resonances}.} \label{warp-d1_5-3}
\end{figure*}

\begin{figure*}
\centering
\includegraphics[width=12.0cm]{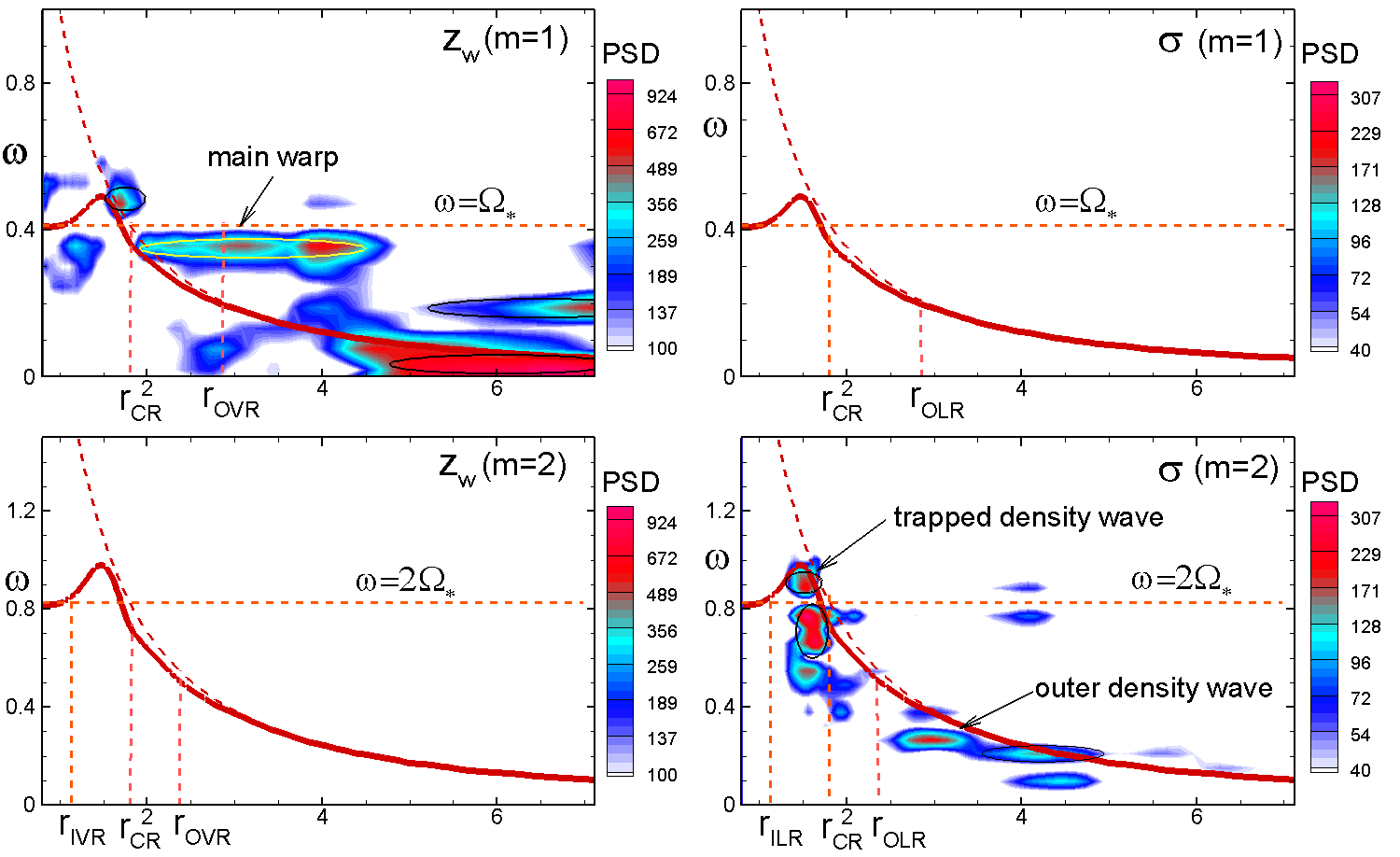}
\caption{Power spectral density (PSD) obtained for waves in model FW$\mu$1.5 inside
$r<7$  of the simulation region (the entire region is $R_{\rm out}=12.1$).
\textit{Left panels:} PSD for one-armed, $m=1$, (top) and two-armed, $m=2$, (bottom)
bending waves. \textit{Right panels:} same, but for density waves. The solid red lines
show the angular velocity distribution in the disc. Dashed lines show angular
frequency of the star and positions of resonances. All frequencies are given in the
non-rotating reference frame. Position of resonances is taken from Tab.
\ref{tab:resonances}.} \label{psd-d1_5-c1_8-2}
\end{figure*}

\subsection{Numerical model and reference values} \label{sec:nummodel-refunits}

We solve the 3D MHD equations with a Godunov-type code in a reference frame rotating
with the star, using the ``cubed sphere" grid. The model has been described earlier in
a series of previous papers (e.g., \citealt{kold02,roma03,roma04}). Hence, we will
describe it only briefly here.

\textit{Initial conditions.} A rotating magnetic star is surrounded by an accretion
disc and a corona. The disc is cold and dense, while the corona is hot and rarefied,
and at the reference point (the inner edge of the disc in the disc plane),
$T_c=100T_d$, and $\rho_c=0.01\rho_d$, where subscripts `d' and `c' denote the disc
and the corona, which are initially in rotational hydrodynamic equilibrium (see e.g.,
\citealt{roma02} for details). The disc is relatively thin, with the half-thickness to
radius ratio $h/r\approx 0.1$.

\textit{Boundary conditions.} At both the inner and outer boundaries, most of the
variables $A_j$ are taken to have free boundary conditions, ${\partial A_j}/{\partial r}=0$.
     On the star (the inner boundary) the magnetic field is frozen onto the surface of the
star. That is, the normal component of the field, $B_n$, is independent of time.
    The other components of the
magnetic field vary. At the outer boundary, matter is not permitted to flow into the
region. The simulation region is usually large enough that the disc has enough mass to
sustain accretion flow for the duration of the simulation. The free boundary
conditions on the hydrodynamic variables at the stellar surface means that accreting
gas can cross the surface of the star without creating a disturbance in the flow. This
also neglects the complex physics of interaction between the accreting gas and the
star, which is expected to produce a strongly non-stationary shock wave in the stellar
atmosphere (e.g., \citealt{kold08}).

\textit{A ``cubed sphere" grid} is used in the simulations. The grid consists of $N_r$
concentric spheres, where each sphere represents an inflated cube. Each of the six
sides of the inflated cube has an $N\times N$ curvilinear grids which represent a
projection of the Cartesian grid onto the sphere. The whole grid consists of $6\times
N_r\times N^2$ cells. The typical grid used in our simulations has $N^2 = 61^2$
angular cells in each block. We use a different number of grid cells in the radial
direction: $N_r=130, 140, 160$.

\textit{Reference values.} The simulations are performed in dimensionless variables
$\widetilde{A}$. To obtain the physical dimensional values $A$, the dimensionless
values $\widetilde{A}$ should be multiplied by the corresponding reference values
$A_0$ as $A=\widetilde{A}A_0$. These reference values include: mass $M_0=M_*$, where
$M_\star$ is the mass of the star; distance $R_0=R_*/0.35$, where $R_*$ is the radius
of the star \footnote{This value for the scale has been taken in the past models
(e.g., Koldoba et al. 2002; Romanova et al. 2002), and now we keep it for consistency
with earlier work.}; and velocity $v_0=(GM/R_0)^{1/2}$. The
 time scale is the period of rotation at $r=R_0$: $P_0=2\pi R_0/v_0$. Reference angular velocity is
$\omega_0=v_0/R_0$, reference frequency is $\nu_0=\omega_0/2\pi$.
      To better resolve the temporal behavior
of the different waves, we record the simulation data every quarter of rotation.  In
the figures we show time $t$ in units of $P_0/4$. Let $\mu_*=B_* R_*^3$ be the
magnetic moment of the star, where $B_*$is the magnetic field at the magnetic equator.
We determine the reference magnetic field $B_0$ and magnetic moment $\mu_0=B_0 R_0^3$
such that $\mu_0=\mu_*/\widetilde\mu$, where $\widetilde\mu$ is the dimensionless
magnetic moment of the star, which is used as a parameter in the code to vary the size
of the magnetosphere. The reference field is
$B_0=\mu_*/(R_0^3\widetilde\mu)$\footnote{subsequent reference values depend on
$\widetilde\mu$}, the reference density is $\rho_0=B_0^2/v_0^2$ and the reference mass
accretion rate is $\dot{M}_0=\rho_0 v_0 R_0^2 = \mu_*^2/(\widetilde{\mu}^2 R_0^4
v_0)$. Reference surface density $\sigma=\rho_0 R_0$. We use in our simulations
$\widetilde\mu=0.5$ for most runs and $\widetilde\mu=1.5$ for sample runs with a
larger magnetosphere. The main reference values are given in Tab. \ref{tab:refval} for
$\widetilde\mu=0.5$.

\begin{figure*}
\centering
\includegraphics[width=16cm]{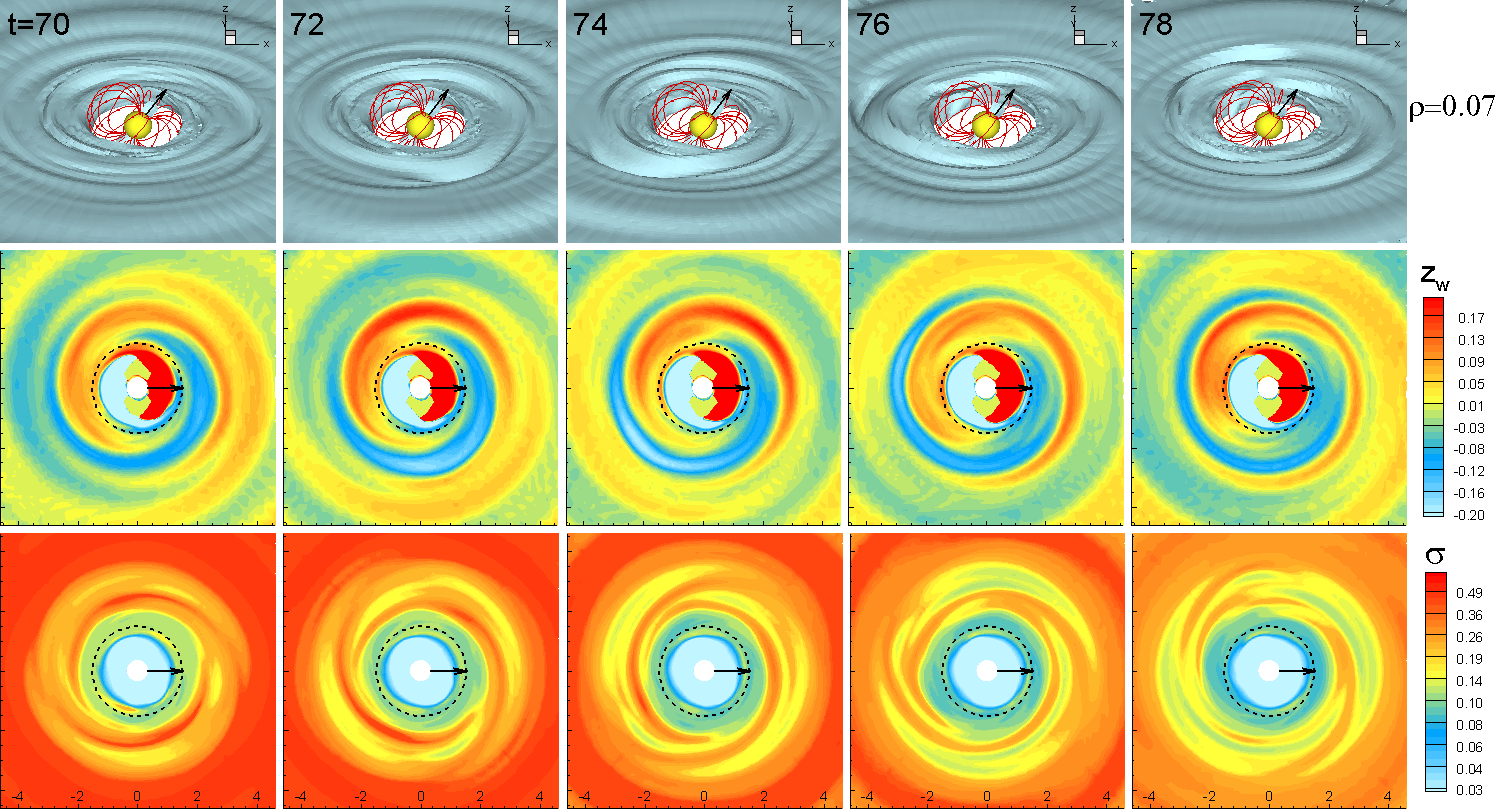}
\caption{Same as in Fig. \ref{warp-d1_5-c1_8-15}, but for a smaller magnetosphere,
$\widetilde\mu=0.5$ (model FW$\mu$0.5).} \label{warp-d0_5-c1_5-15}
\end{figure*}

\begin{figure*}
\centering
\includegraphics[width=16cm]{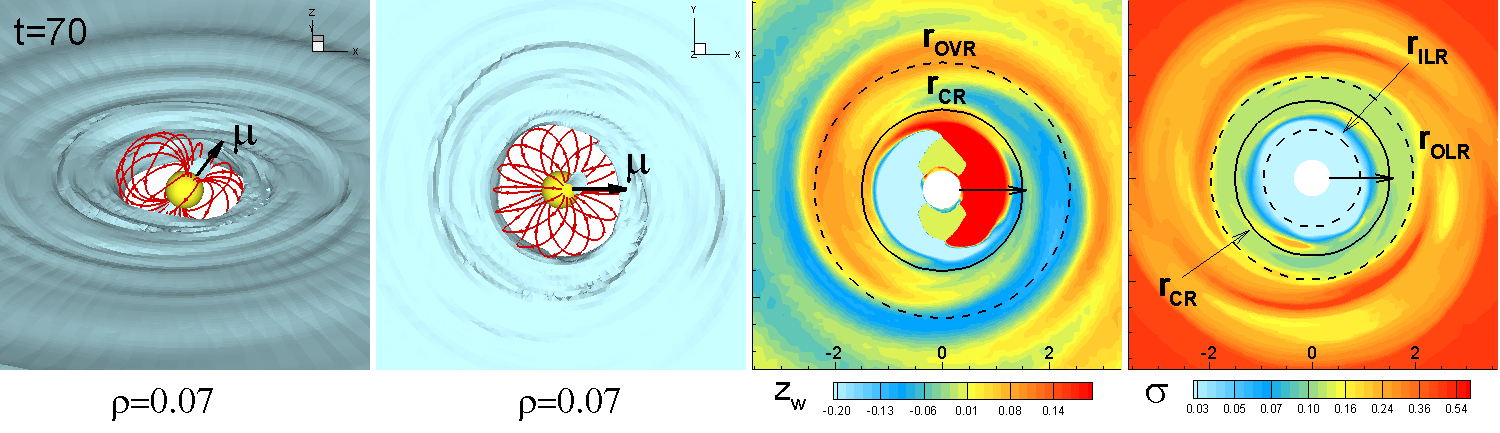}
\caption{Same as in Fig. \ref{warp-d1_5-3}, but for a smaller magnetosphere,
$\widetilde\mu=0.5$ (model FW$\mu$0.5).} \label{warp-d0_5-3}
\end{figure*}

\textit{Numerical method and code:} In our Godunov scheme, we use the solver described
by \citet{powell99}.
      We use a coordinate system rotating with the star.
  We split the magnetic field into a vacuum dipole component  and a component
calculated in equations: $\bf B=\bf B_d+\bf B'$ \citep{tanaka94}.
    We solve the full set of equations for the ideal magnetohydrodynamics (e.g., Koldoba et al. 2002).
 \textbf{Instead of using the full energy equation, we use an equation for
the entropy\footnote{This is reasonable because we do not expect the formation of
shocks.} in the form: $\partial K(S)/\partial t + (\textbf{u}\cdot \bigtriangledown)
K(S)=0$, where $K(S)=p/\rho^\gamma$ is a function of  the specific entropy $S$
\footnote{The entropy can be derived in the standard form, $S =
\log(p/\rho^\gamma)/(\gamma-1)$.}. However, to characterize the thermodynamic
properties of the disc, we use the temperature derived from the the ideal gas
equation, $T = p/\rho {\mathcal R}$.}

Viscosity is included in the code, with the $\alpha-$prescription for the viscosity
coefficient \citep{shakura73}. Viscosity supports slow, quasi-steady accretion of
matter towards the star during  long time intervals.
  The viscosity is
nonzero only inside the disc, that is, above a certain threshold density (which is
$\rho = 0.2$ in our dimensionless units). We use small value $\alpha=0.02$ in most of
the simulation runs, but take even smaller, as well as larger values in the test runs.

A typical simulation run lasts about $50 P_0$. This time is sufficient to observe and
analyze most frequencies in the disc. To analyze global, low-frequency modes in the
disc, we use a smaller simulation region.

We consider only non-relativistic discs.
  For discs around neutron stars the relativistic effects can be
important and can significantly modify the waves (e.g., \citealt{nowak91,kato98}).
However, test simulations of waves performed in the weakly relativistic case show
results similar to those of the non-relativistic case.

\begin{figure*}
\centering
\includegraphics[width=12.0cm]{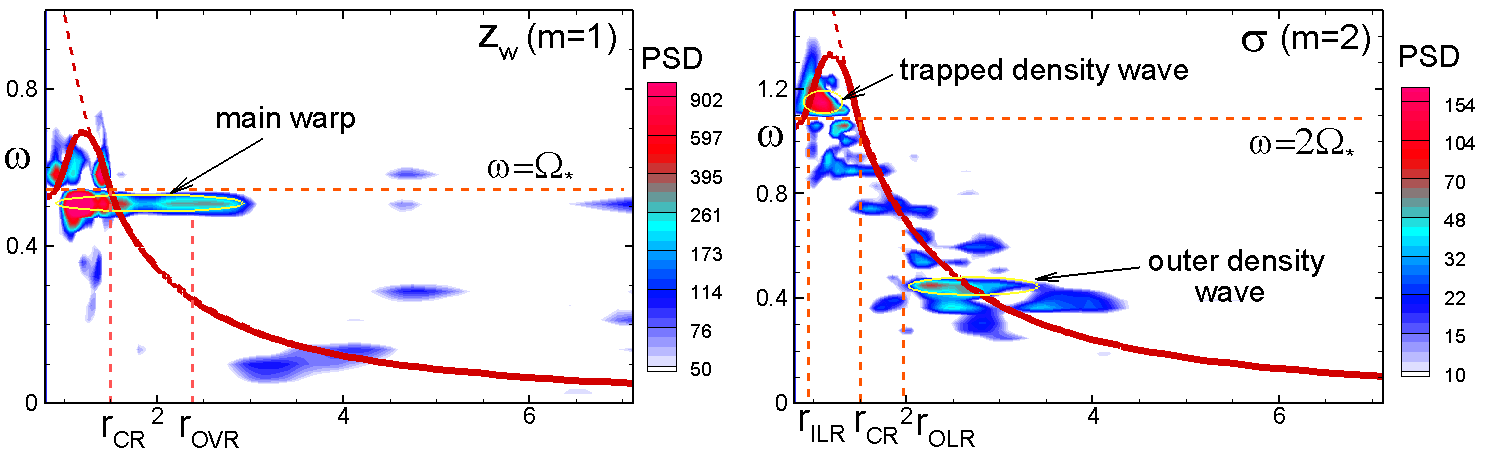}
\caption{Power spectral density obtained for waves in model FW$\mu$0.5 inside $r<7$ of
the simulation region. \textit{Left panel:} PSD for one-armed, $m=1$, bending wave.
\textit{Right panel:} PSD for two-armed, $m=2$, density waves.}
\label{psd-d0_5-c1_5-2}
\end{figure*}

\subsection{Analysis of waves in the disc}
\label{sec:theory-analysis of waves}

Here, we describe  the methods used  to analyze the different waves  observed in our
3D MHD simulations.

\subsubsection{Visualization of density and bending waves}

To reveal the position and amplitude of density and bending waves, we separate the
disc from the corona using one of the density levels, $\rho_0=0.03$, and then
integrate through the disc in the $z-$direction.

To analyze the density waves, we calculate the  surface density at each point of the
disc for different moments of time:
\begin{equation}
\sigma=\sigma(t,r,\phi)= \int dz\rho  ~, \label{eq:sigma}
\end{equation}
where $r$ is the radius in the equatorial plane and $\phi$ is the azimuthal angle. As
a result, we obtain the two-dimensional distribution of density waves.

To analyze the bending waves, we calculate the vertical displacement of the center of
mass for each point of the disc:
\begin{equation}
 z_w(t,r,\phi) = \frac{\int dz\rho z   }{\int dz\rho }~ ,
 \label{eq:zmass}
\end{equation}
and obtain a two-dimensional distribution of bending waves for different times.

Typically, most of the matter in the disc has a density of $\rho = 0.1$ or higher. We
chose a smaller value for  $\rho_0$ in order to be sure that we do not miss the
regions between spiral waves, or other locations with low density.

We use the time-dependent distribution of $\sigma(t,r,\phi)$ and $z_w(t,r,\phi)$ for
visualization of density and bending waves, and also for spectral analysis described
in the next section.

\subsubsection{Spectral analysis of disc oscillations} \label{sec:frequency-analysis}

Here, we analyze the distribution of the angular frequencies of waves in the disc. We
consider the variables $ z_w(t,r,\phi)$ and $\sigma(t,r,\phi)$ and denote them as
 $f_j(t,r,\phi)$, $j=1,~2$.
     This is a function of time $t$, and the two
spatial coordinates,
 $r$ (radius in the equatorial plane) and $\phi$ (azimuthal angle).
      Our simulations use a coordinate system rotating with the star, and the functions $f$
      are calculated in this reference frame.
       At a fixed position ($r,~\phi)$ in the  equatorial plane
 the function $f$ is quasi-periodic, because there are always some
 inhomogeneities in the disc which move faster or slower than the coordinate
 system.
      They cycle around the rotation axis of the disc. The function $f$ is determined in some interval of
time, $t_1<t<t_2$. It is not periodic, but rather oscillates around some value, which
by itself also varies in time. In addition, it rotates relative to a distant observer
with the angular velocity of the star, $\Omega_*$.
    Our objective  is to identify the most
significant frequencies as seen by a distant observer.
    Also,  we want
 to determine the distribution of waves in the disc.

As a first step of analysis, we ``clean"  the function $f$.  The goal of this cleaning
is to exclude from consideration parts of the data which are not needed for frequency
analysis, for example, the average density, or regular variations of the function $f$.
      In general $f(t_1) \ne f(t_2)$, and in the observer's
frame, $f^{\rm obs}(t_1) \ne f^{\rm obs}(t_2)$.
    We consider the function which has been obtained as a result of this cleaning
    as a function determined for the whole
time-interval $t_2-t_1$, with period $T=t_2-t_1$. In order to do this, we subtract
from $f(t)$ the linear function:
\begin{equation}
f(t) \to f(t) - \frac{f(t_2) - f(t_1)}{t_2-t_1} (t - t_1),
\end{equation}
and then subtract from the resulting function the  time-averaged value
\begin{equation}
f(t) \to f(t) - \langle f\rangle.
\end{equation}
In the next step, we select the azimuthal modes of the function $f$. The function
$f^{\rm obs}$  (which is in the observer's frame) is connected to function $f$ in the
star's frame by the relationship:  $f^{\rm obs}(t,r,\phi) = f(t,r,\phi - \Omega_* t)$.

We use the Fourier expansion of $f^{\rm obs}$:
\begin{equation}
f^{\rm obs}(t,r,\phi)= \sum f_m^{\rm obs}(t,r)\exp(im\phi)~,
\label{eq:f1}
\end{equation}
where
\begin{multline}
f_m^{\rm obs}(t,r) = \frac{1}{2\pi} \int d\phi f^{\rm obs}(t,r,\phi) \exp(-i m \phi) \\
= \frac{1}{2 \pi} \int d\phi f(t,r,\phi- \Omega_*t) \exp(-i m \phi) \\ = \frac{1}{2
\pi} \int d\chi f(t,r,\chi) \exp[-i m (\chi + \Omega_* t) ] \\ = f_m(t,r) \exp(-i m
\Omega_* t ). \label{eq:f2}
\end{multline}

In the next step, we analyze the function  $f_m^{\rm obs} = f_m \exp(- i m \Omega_*
t)$. Functions $f_m$ and $f_m^{\rm obs}$ are also determined in the interval $t_1\leq
t\leq t_2$. Subsequently, we perform a spectral analysis of $f_m$, and interpret the
term $\exp(-i m \Omega_* t )$ as the displacement of the frequency by  the value $m
\Omega_*$.

   Next, we perform the time Fourier transform  of $f_m(t,r)$.
We consider  $f_m$ as a periodic function of time with period $T=t_2-t_1$. In this
case, the angular frequency has a discrete set of values with step $\delta \omega =
2\pi/T$. The coefficients of the Fourier transformation are (here we neglect the
numbers corresponding to frequencies):
\begin{multline}
{\hat f}_m^{\rm obs}(\omega,r) = \frac{1}{T} \int dt f_m^{\rm obs} (t,r) \exp(i \omega
t)
 \\ = \frac{1}{T} \int dt f_m(t,r) \exp(i \omega t - i m \Omega_* t)
 =
 {\hat f_m}(\tilde\omega,r), \label{eq:f3}
\end{multline}
where, $\tilde\omega\equiv \omega - m\Omega_*$.
     Hence, the Fourier coefficients of the function  $f_m(t,r)$, determined in the
rotating coordinate system, are calculated for frequencies considered in the rotating
coordinate system. When we transfer to the observer's coordinate system, all
frequencies are misplaced:  $\omega = \tilde\omega + m \Omega_*$, but the Fourier
coefficients do not change.

\begin{figure*}
\centering
\includegraphics[width=11.0cm]{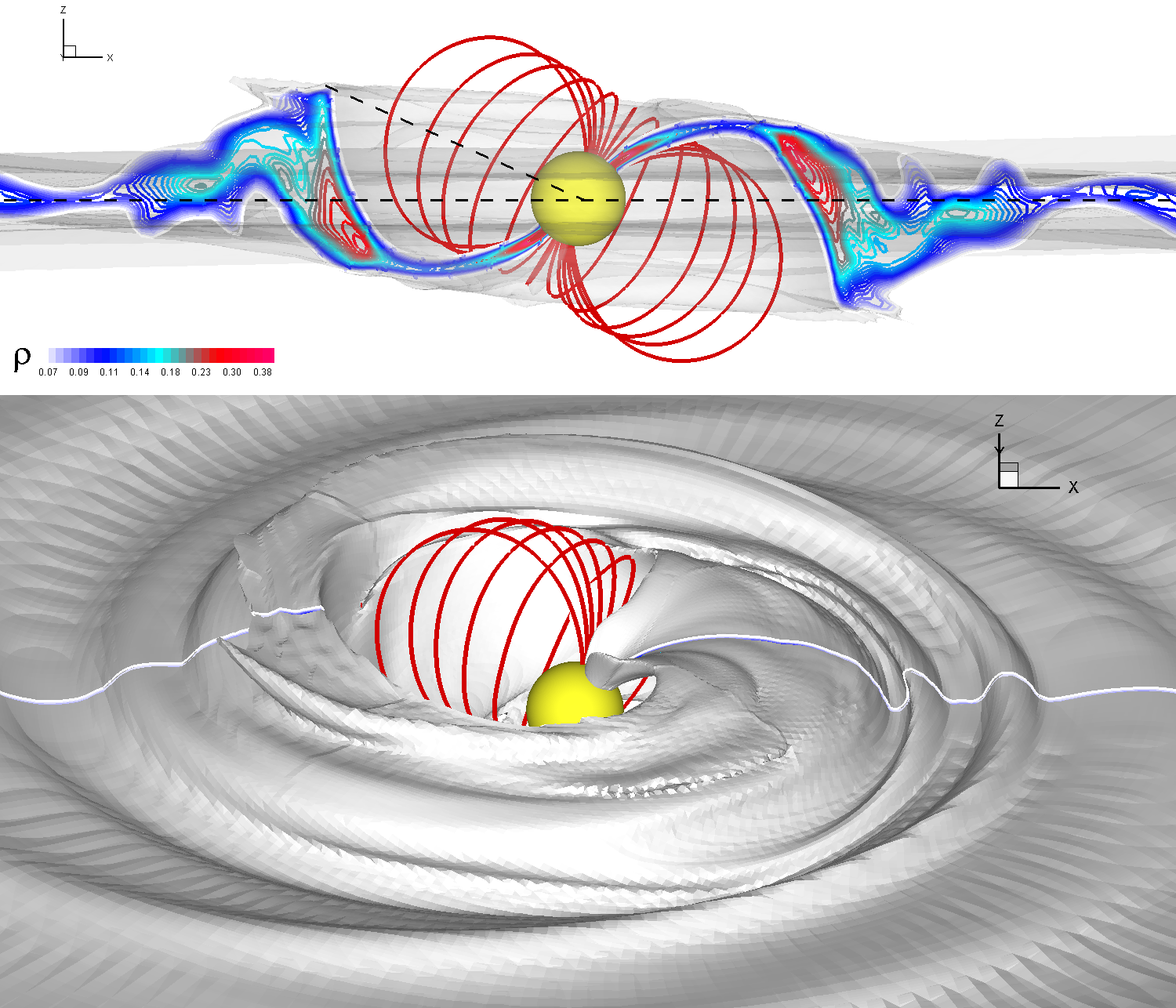}
\caption{Analysis of the height of the warp in model FW$\mu$1.5 at $t=80$. \textit{Top
panel:} Cross-section of the density distribution in the $xz-$plane. Sample magnetic
field lines show the location of the magnetosphere. \textit{Bottom panel:} 3D view of
the magnetosphere and the warp.  The white line shows the location of the vertical
slice shown in top panel. } \label{height-warp}
\end{figure*}

To characterize the spectral properties of the function $f_m(t,r)$, we use the value
of the Power Spectral Density (PSD):

\begin{equation}
{\rm PSD}=|{\hat f}_m^{\rm obs}(\omega,r)|^2 . \label{eq:f4}
\end{equation}

\noindent This value characterizes the power in the signal with frequency  $\omega$.
Here, we analyze one-armed  ($m=1$) and two-armed  ($m=2$) modes for density and
bending waves.

\section{Waves observed in 3D MHD simulations}
\label{sec:results-main}

We performed multiple simulation runs for different parameters of the star at fixed
parameters of the disc (see Sec. \ref{sec:simulation runs} and Table
\ref{tab:models}). The results strongly depend on the relative positions of the
magnetospheric radius, $r_m$ \footnote{$r_m$ is a radius, where the matter stress in
the disc equals the magnetic stress in the magnetosphere} and the corotation radius,
$r_{cr}$.  We can separate our results into two main groups: (1) The inner disc
rotates with angular velocity close to that of the star (that is, $r_m\approx
r_{cr}$).
     In this case, the rotating magnetosphere excites a high-amplitude bending wave (a
     warp),
      which  rotates with the angular velocity of the star (see Sec. \ref{sec:fast warp}).
      (2) The inner disc rotates faster than the star (that is, $r_m <
r_{cr}$). In these models, the high-amplitude (corotating with the star) warp does not
form (see Sec. \ref{sec:slow warp section}).


\subsection{Description of simulation runs}
\label{sec:simulation runs}

 We use as a base case a model with the following parameters: the magnetic moment $\widetilde\mu=0.5$ (in
dimensionless units, see Sec. \ref{sec:nummodel-refunits}), so that the magnetospheric
radius obtained from simulations is $r_m\approx 1.3$ (or, in stellar radii,
$r_m\approx 3.7 R_*$). Such a magnetosphere is sufficiently large to excite waves in
the disc.
    At the same time the
simulation runs have a reasonable duration (on a supercomputer running at 168
processors), so that we were able to run multiple cases with different parameters.
      The star rotates with
angular velocity $\Omega_*=0.54$ (in dimensionless units).
    The corresponding corotation
radius is  $r_{cr}=1.5$ is only slightly (1.15 times) larger than the magnetospheric
radius.
    We take the angle between the star's magnetic moment
 and the spin axis (the misalignment angle) to be $\theta=30^\circ$.
    The maximum (outer) radius of the simulation
region is $R_{out}=12.1$ (in our dimensionless units), or $R_{out}=12.1/0.35= 34.5
R_*$ (in radii of the star). We use the described model as a base and call it
FW$\mu$0.5. Abbreviation ``FW" indicates a ``Fast Warp" because in all cases where
$r_m\approx r_{cr}$, a strong bending wave (a warp) forms and rotates with the angular
velocity of the star, which is fast, compared to other cases.

We varied different parameters, taking model FW$\mu$0.5 as a base model. In model
FW$\mu$1.5 we show an example of a larger magnetosphere, $\widetilde\mu=1.5$, where
similar waves are excited but with larger amplitudes (see Sec. \ref{sec:model-mu1.5}).
   We discuss the warp amplitude in this model in Sec. \ref{sec:amplitude of warp}.

In models FW$\theta$5, FW$\theta$15, FW$\theta$45, FW$\theta$60 and FW$\theta$90 we
investigate excitation of waves  for different misalignment angles $\theta$ and
compare them with the base case FW$\mu$0.5, which corresponds to $\theta=30^\circ$
(see Sec. \ref{sec:tilts}).

 In models FW$\alpha$0.00,
FW$\alpha$0.04, FW$\alpha$0.06, FW$\alpha$0.06 and FW$\alpha$0.08, we investigate the
possible dependence of warp properties on the  $\alpha-$coefficient of viscosity (see
Sec. \ref{sec:viscosity}).

\begin{figure*}
\centering
\includegraphics[width=12.0cm]{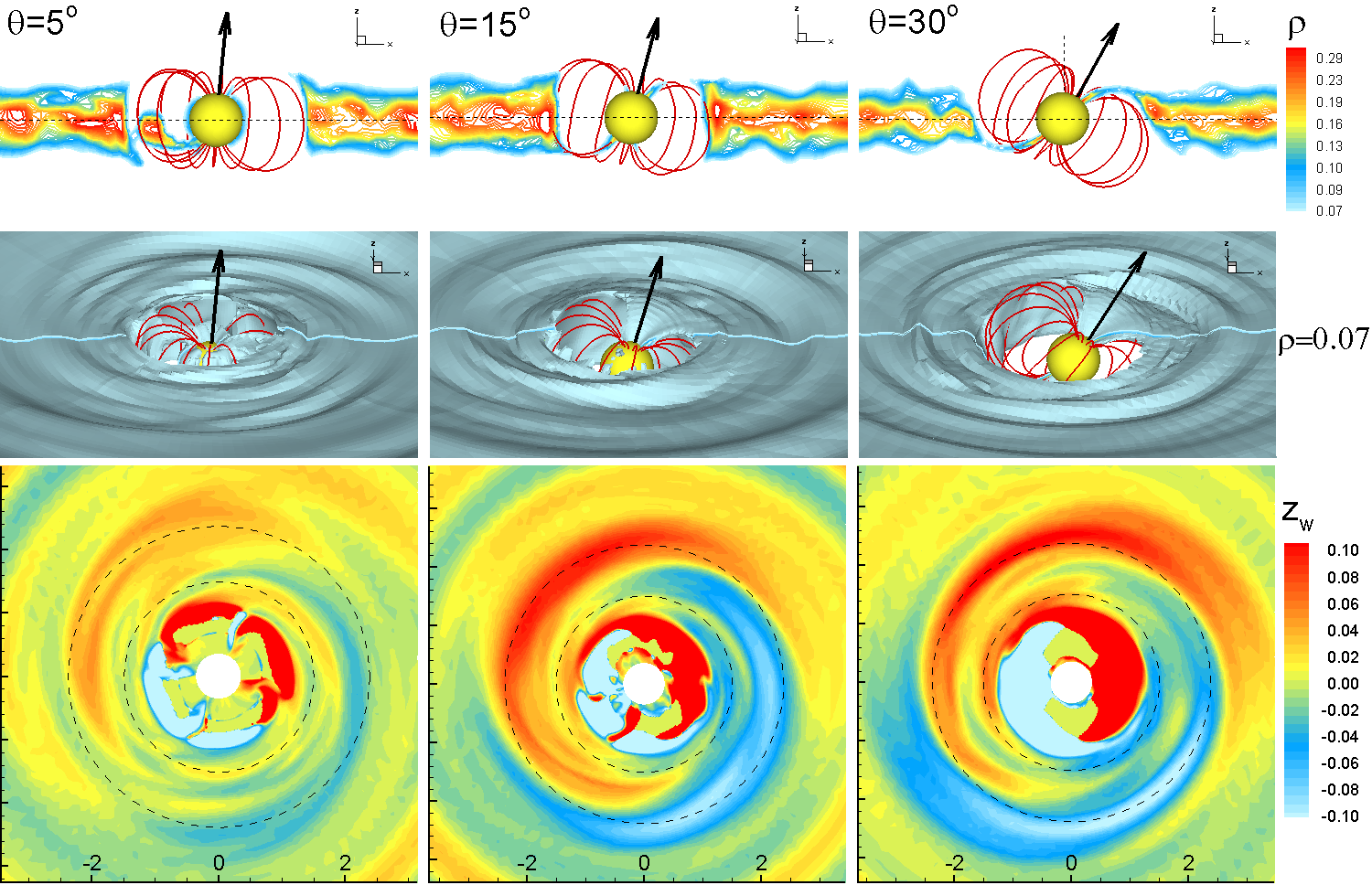}
\caption{Analysis of bending waves in the disc in cases where the magnetic moment of
the star is tilted at angles $\theta=5^\circ, 15^\circ$ and $30^\circ$ (models
FW$\theta5$, FW$\theta15$, and FW$\theta30$). \textit{Top panels:} $xz$-slices of
density distribution and sample magnetic field lines.  \textit{Middle panels:} 3D view
of matter flow at one density level, $\rho=0.07$. \textit{Bottom panels:}
 $z-$averaged height of the local disc's center of mass
above or below the equatorial plane, $z_w$.  Dashed circles show the location of
corotation and vertical resonances: $r_{\rm CR}=1.5$, $r_{\rm OVR}=2.38$.}
\label{warp-theta-5-15-30}
\end{figure*}

\begin{figure*}
\centering
\includegraphics[width=12.0cm]{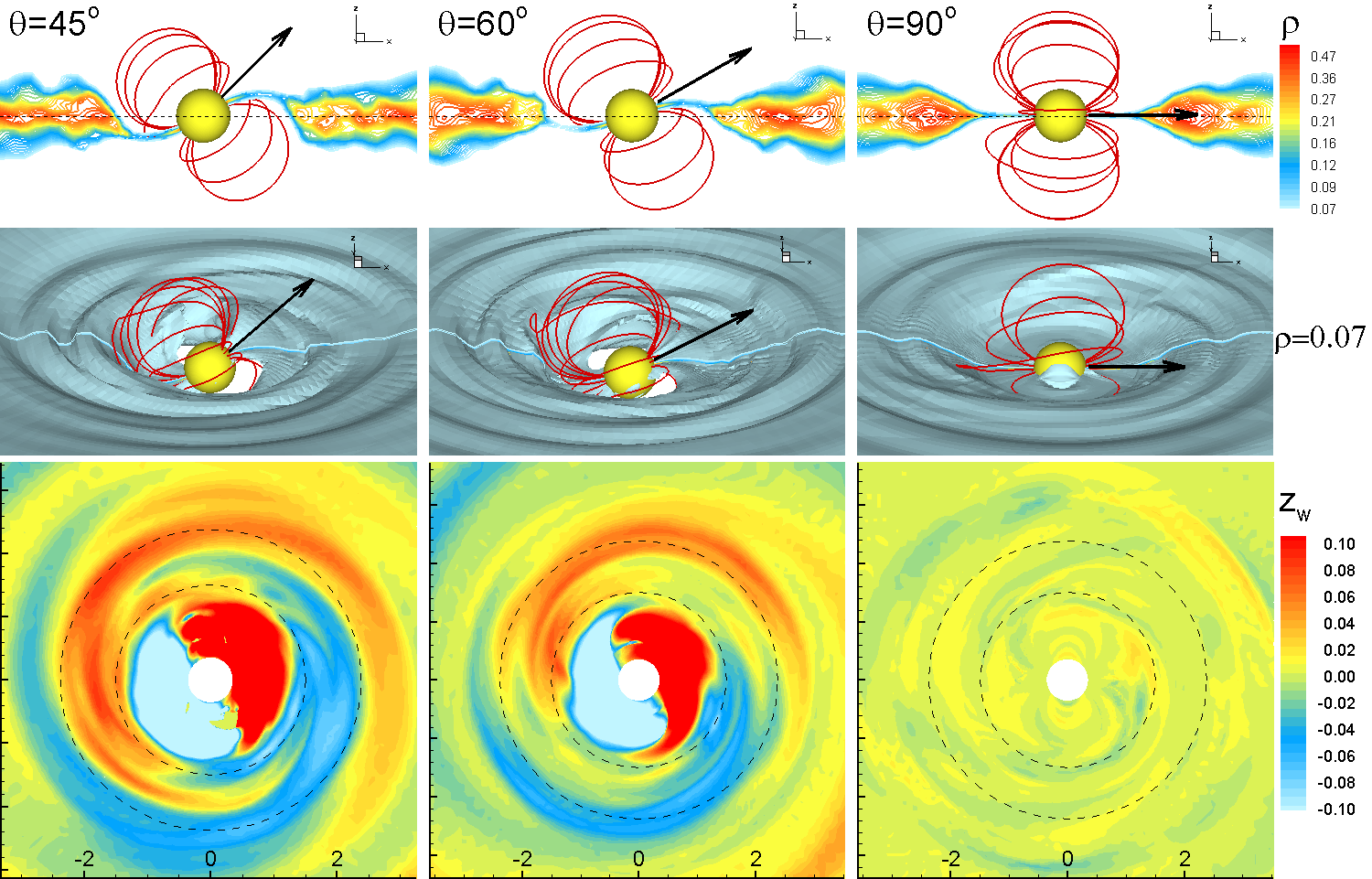}
\caption{Same as in Fig. \ref{warp-theta-5-15-30}, but for misalignment angles of the
dipole $\theta=45^\circ, 60^\circ$ and $90^\circ$ (models FW$\theta45$, FW$\theta60$,
and FW$\theta90$).} \label{warp-theta-45-60-90}
\end{figure*}

In another set of simulation runs we drop the condition $r_m\approx r_{cr}$ and
decrease the angular velocity of the star, $\Omega_*$  (thereby increasing the
corotation radius $r_{cr}$).
     We observed that the strong magnetospheric
warp disappears but other types of waves become stronger.
    In some cases, a slow warp
is observed (see models SWcor1.8, SWcor3 and SWcor5).
     Experiments with different
sizes of the simulation region show that this warp appears only in models with
relatively small simulation region, where free disc oscillations are excited more
rapidly. We demonstrate excitation of free disc oscillations using a smaller
simulation region, $R_{out}=9.4=26.9 R_*$ (model SRcor3).
     In another set of
simulation runs with a large region, $R_{out}=20.2=57.7 R_*$, a warp is not present
(models LRcor1.8, LRcor3 and LRcor5).
     However, apart from the main magnetospheric
warp, which can be present or not, other types of bending and density waves were
observed in all simulation runs.

We calculated the position of resonances for all models using the theoretical formulae
for the thin disc (see Sec. \ref{sec:theory}) which are given in Table
\ref{tab:resonances}.  We compare the position of waves obtained in 3D simulations,
with the position of resonances.

\subsection{Waves and Fast Warps in Cases where $r_m\approx r_{cr}$}
\label{sec:fast warp}

Here, we investigate the case of  rapidly-rotating stars, where the corotation radius
is approximately equal to the magnetospheric radius, $r_m\approx r_{cr}$. This case is
interesting because it approximately corresponds to the rotational equilibrium state
which is the most probable state in the life of accreting magnetized stars (e.g.,
\citealt{long05}).

We performed two simulation runs (models FW$\mu$0.5 and FW$\mu$1.5) for stars with
different sizes of the magnetosphere ($\widetilde\mu=0.5$ and $\widetilde\mu=1.5$) and
matching corotation radii $r_{cr}=1.5$ and $r_{cr}=1.8$ (so that $r_m\approx r_{cr}$).
The dipole magnetosphere is tilted at $\theta=30^\circ$.
      We observed that in both cases, a large warp forms at
the corotation radius, and it rotates with the angular velocity of the star. We use
these two cases to demonstrate the formation of magnetospheric warps and to show their
main properties.

\subsubsection{Formation of fast warp and other waves in the case of a large magnetosphere (Model FW$\mu$1.5)}
\label{sec:model-mu1.5}

First, we investigate the case of large magnetospheres (model FW$\mu$1.5), where the
warp is larger than in the base case. We observed from simulations that the
magnetosphere truncated the disc at the distance of $r_m\approx 1.55$ from the star.
This radius $r_m$ is $1.16$ times smaller than $r_{cr}=1.8$ and therefore $r_m\approx
r_{cr}$. Fig. \ref{warp-1} shows that matter of the inner disc has been lifted above
the disc and flows to the star in two ordered funnel streams (see
\citealt{roma03,roma04}). It can also be seen that the inner parts of the disc have a
bent structure. In particular,  the far left regions of the disc are lifted above the
rest of the disc, forming a warp.

Fig. \ref{warp-d1_5-c1_8-15} (top panels) shows three-dimensional views of the inner
disc density distribution (one of the density levels) at different moments in time
(plots are shown in the coordinate system rotating with the star).
    One can see that the location of the high-amplitude
bending wave (a warp), seen in Fig. \ref{warp-1}, does not change with time (in the
coordinate system rotating with the star), and thus the warp corotates with the star
and its magnetosphere.
    The middle panels show the
height of the center of mass of the disc $z_w$ (see eq. \ref{eq:zmass}) above or below
the equatorial plane. It can be seen that the location of the warp coincides with the
inner, high-amplitude part of the bending wave, which also corotates with the star
(red color).
    The bending wave is evident on the opposite side of the disc (blue
color). Thus, we can see that the bending wave represents a one-armed ($m=1$) feature
which propagates to larger distances.
    The bottom panels show the $z-$averaged density distribution in the accretion disc,
$\sigma$ (see eq. \ref{eq:sigma}). One can see that the two-armed ($m=2$) trailing
density wave ($n=0$)  forms at the corotation radius and propagates to larger
distances.
   The formation of the warp which corotates with the magnetosphere
is in agreement with theoretical predictions \citep{terquem00}.
    The warp does not precess  (as discussed by \citealt{lai99}),
because in our simulations the rotational axis of the dipole is aligned with the
rotational axis of the disc (like in \citealt{terquem00}).

The left two panels in Fig. \ref{warp-d1_5-3}  show an expanded view of the warp shown
in Fig. \ref{warp-d1_5-c1_8-15}, both in side and face projections.
    The right two panels
show the locations of resonances calculated theoretically for a thin Keplerian disc
(see Table \ref{tab:resonances} and Appendix \ref{sec:Appendix-allwaves-modes}). One
can see that the amplitude of the bending wave increases at the corotation radius,
stays high at the Outer Vertical Resonance (OVR) and gradually decreases  at larger
radii, where it forms a trailing spiral wave.
     According to the theory (e.g., \citealt{kato98,zhang06}, see Sec.
\ref{sec:Appendix-warping-modes} of the Appendix A) a bending wave is expected to have
its maximum amplitude at and around vertical resonance (see Fig. \ref{Appfig:allwarps}
of the Appendix, 3rd panel from the left).
    Fig. \ref{warp-d1_5-3} shows that in our simulations,
the warp has its maximum amplitude between CR and  OVR.

\begin{figure*}
\centering
\includegraphics[width=15.0cm]{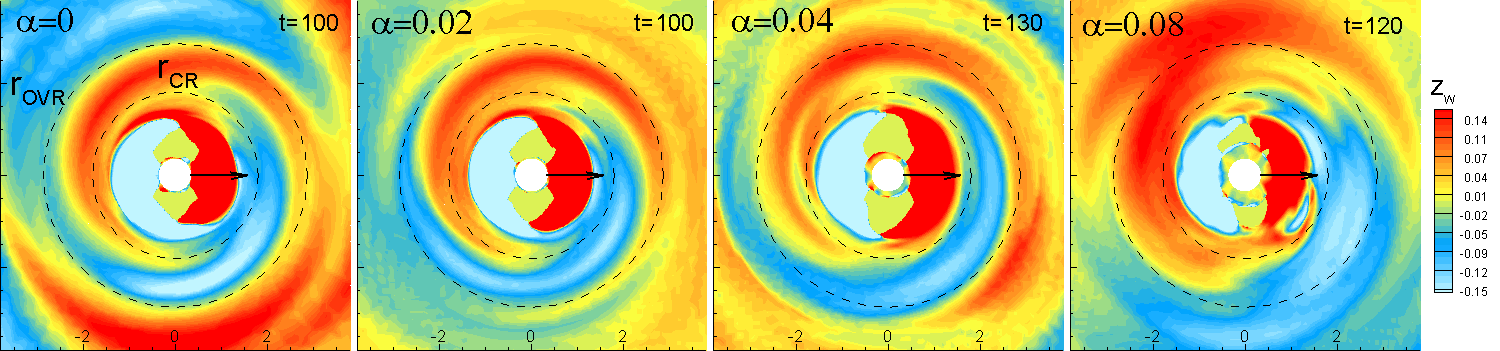}
\caption{Warps observed in simulations where the viscosity parameter $\alpha$ varies
from $\alpha=0.0$ to $\alpha=0.08$ (models FW$\alpha$0.00 - FW$\alpha$0.08 ).}
\label{warp-visc-4}
\end{figure*}

 The shape of the  bending wave observed in our simulations is similar to
that derived from the theoretical analysis (see right panel of Fig.
\ref{Appfig:allwarps} in Sec. \ref{sec:Appendix-warping-modes}). The right-hand panel
shows the distribution of density waves, and the locations of the resonances.

      According to the theory (e.g.,
\citealt{zhang06}), one-armed ($m=1$) in-plane ($n=0$) density waves can propagate to
the exterior of OLR, while two-armed ($m=2$) waves can propagate to the exterior of
OLR and to the interior of ILR.
     In our case, there are clear, two-armed density waves which
propagate at $r>r_{\rm OLR}$, which is in agreement with the theory.
     The ILR is located inside the magnetosphere and therefore, there
is no disc at $r<r_{\rm ILR}$. It is possible that these density waves are excited
directly by the magnetic forcing on the inner parts of the disc. Alternatively, the
magnetic force may excite predominantly bending waves, which in turn compress matter
and generate density waves. A one-to-one comparison between bending and density waves
shows some similarity in the pattern between them, so that this is a possibility.

We apply the spectral analysis described in Sec. \ref{sec:frequency-analysis} to this
model (FW$\mu$1.5).
    The top left panel of Fig. \ref{psd-d1_5-c1_8-2} shows PSD for the one-armed
($m=1$) bending ($n=1$) waves.
    It can be seen that PSD has a clear feature corresponding
to a warp, which rotates rigidly with the angular velocity of the star.
     This feature
starts at the corotation radius and extends outward  to  a distance of $r\approx 2.7
r_m$.
    There are also lower-frequency bending waves, which are at frequencies above or below Keplerian.
    The higher-frequency waves may represent a coupling between
    free bending oscillations which have approximately the local Keplerian velocity
 and bending waves excited by the star.
Waves of the lowest frequency may represent a coupling between Keplerian frequency and
the frequency of free, bending oscillations of the disc.
    Analysis of the two-armed ($m=2$) bending waves shows
that these waves are much weaker with a PSD negligibly small compared with the
one-armed bending wave.

We also calculated the PSD for one-armed and two-armed density waves, and observed
that the PSD of a two-armed wave is much stronger.
      Fig. \ref{psd-d1_5-c1_8-2} (bottom right panel) shows that for
radii $r_{\rm OLR}<r<5$, there is a feature with angular frequency $\omega\approx
0.25-0.3$ which corresponds to the density waves observed in Fig.
\ref{warp-d1_5-c1_8-15} (bottom panels). This frequency corresponds to the frequency
of both waves.  In addition, PSD is high for several frequencies at $r<r_{\rm CR}$.
One of these frequencies (the highest) may correspond to
 trapped waves (e.g., \citealt{love09})(see Sec. \ref{sec:trapped waves}).
 These waves appear where the
angular velocity of the disc decreases to match the angular velocity of the star for
the case where the magnetosphere rotates slower than the inner disc. In this model,
the corotation radius is only slightly larger than the magnetospheric radius.
Nevertheless, there is a maximum in the angular velocity distribution (solid red line
in Fig. \ref{psd-d1_5-c1_8-2}) where trapped waves can form.

Fig. \ref{psd-d1_5-c1_8-2} (right bottom panel) shows a
 candidate trapped wave, which is located interior to the peak in angular velocity distribution at $r\approx 1.5$
and its angular velocity $\omega=0.9$ which is slightly higher than $2\Omega_*\approx
0.83$. The non-Keplerian dependencies of $\kappa$ and  $\Omega$ are fully accounted
for in \citet{love09}.

     There is another density wave at about the same distance, $r\approx 1.5-1.6$  but at a
lower frequency, $\omega\approx 0.7-0.8$. This may be a wave excited by magnetic force
at the disc-magnetosphere boundary, but it weakens at the corotation radius. There is
also a wave at frequency $2\Omega_*$ with smaller PSD. This wave may be excited by the
warp in the disc. One-armed density waves have negligibly small PSD (right top panel).

It is interesting to note that the PSD for density waves is very different from that
for bending waves. This supports the first hypothesis discussed above that bending and
density waves are excited independently by the tilted magnetosphere.



\subsubsection{Formation of fast warp and other waves in the case of a smaller magnetosphere (FW$\mu$0.5 - base model)}\label{sec:model-mu0.5}

We calculated another similar case, but for a star with a smaller magnetic moment,
$\tilde\mu=0.5$.
     In this case the disc is truncated at a smaller radius,
$r_m\approx 1.2$. We took a smaller corotation radius, $r_{cr}=1.5$, in order to have
$r_m\approx r_{cr}$ as in model FW$\mu$1.5 (here, we obtain slightly larger value,
$r_{cr}/r_m\approx 1.25$ compared with the model FW$\mu$1.5, where $r_{cr}/r_m\approx
1.15$).
    We observed that this case is similar to model FW$\mu$1.5,
however all scales are smaller because of the smaller size of the magnetosphere, which
excites the waves.

Fig. \ref{warp-d0_5-c1_5-15} (top panels) show a 3D view of bending waves and the
inner warp. The middle panels show that a bending wave forms at the corotation radius
and propagates outward.
     The  inner part of the bending wave (a warp) is located at
the same position indicating that the warp corotates with the star.
    The bottom panels show the formation of two-armed density waves, similar to model
    FW$\mu$1.5, but at smaller scale.

Fig. \ref{warp-d0_5-3} (left two panels) shows expanded views of the warp in two
projections. The two right panels show the positions of bending and density waves
relative to resonances.
    One can see that the warp part of the bending wave is located between the
corotation and the vertical resonances, with maximum amplitude at the vertical
resonance.  The right-hand panel shows that density waves are excited at the Lindblad
resonance but are damped rapidly at $r\approx 3.4$, which approximately corresponds to
the radial extent of the warp. In addition, one can see density enhancements between
the closed magnetosphere and the corotation radius. These are trapped density waves.

Fig. \ref{psd-d0_5-c1_5-2} (left panel) shows that in the PSD plot the main feature
corresponds to the warp which rotates with the angular velocity of the star and
spreads out to $r\approx 3$, which is  $2.5$ radii of closed magnetosphere.

The right-hand panel of Fig. \ref{psd-d0_5-c1_5-2} shows that the two-armed outer
density wave is located at $r_{\rm OLR}<r\lesssim 4$. The angular frequency (for both
waves) is $\omega\approx 0.4$. PSD also shows a two-armed trapped density wave with
angular frequency $\omega\approx 1.14$ which is located at the inner disc, with
$r\approx 1.15$. The frequency of this wave is located below the maximum in the disc
angular velocity distribution.

Thus, the warp and other features in model FW$\mu$0.5 are similar to those of model
FW$\mu$1.5, which has a larger magnetosphere.
      In both cases, the warp forms near the
magnetosphere and corotates with the star.
   The warp extends out to $r \approx
(2.5-2.7) r_m$ and rotates rigidly in this region.
     Below, we use a model with a
larger magnetosphere, FW$\mu$1.5, to determine the amplitude of the warp. In all other
models, we use model FW$\mu$0.5 as a base case. For this model, the simulation runs
are not as long, and we can investigate the model at a number of different parameters
(see Table \ref{tab:models}).

\begin{figure*}
\centering
\includegraphics[width=16.0cm]{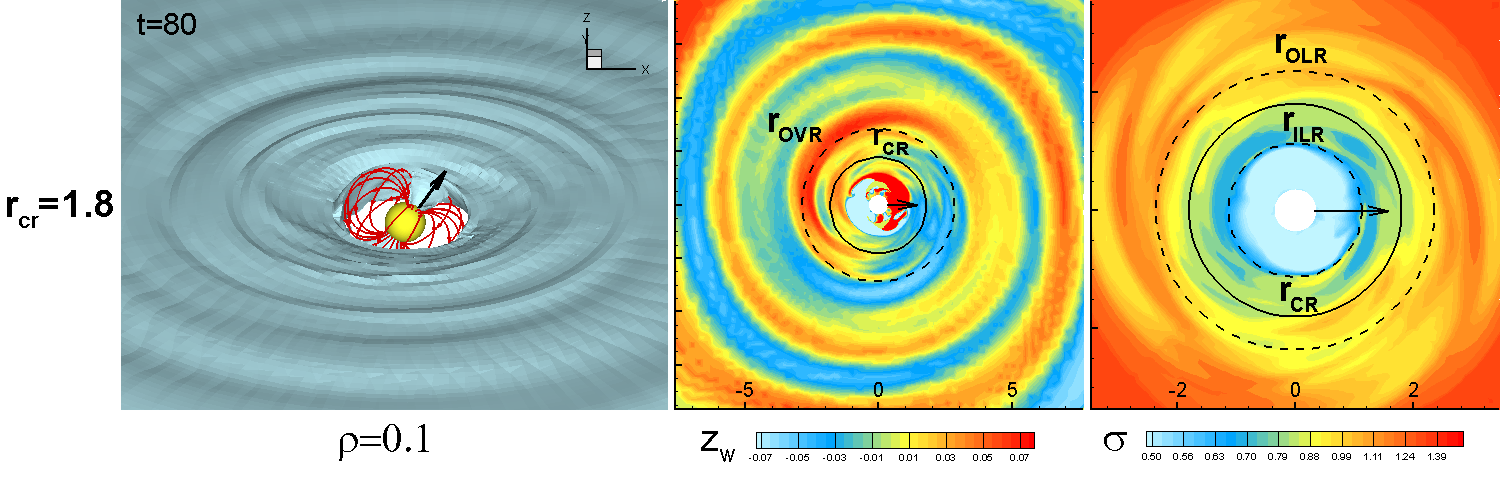}
\includegraphics[width=16.0cm]{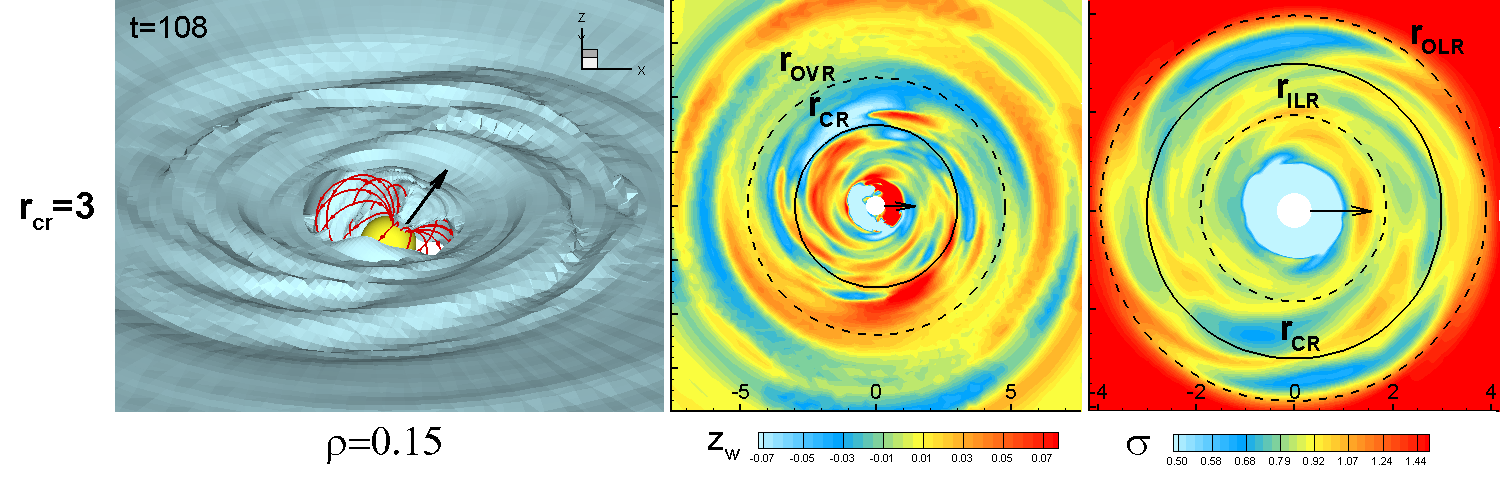}
\includegraphics[width=16.0cm]{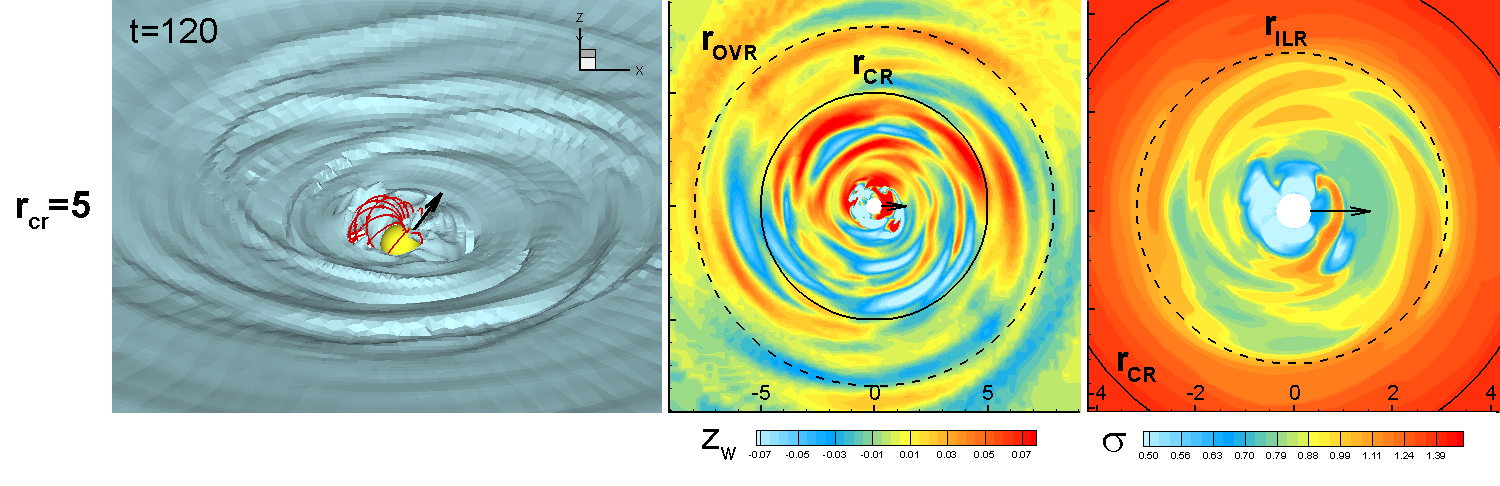}
\caption{Waves observed in models with large simulation region ($R_{out}=20.2=57.7
R_*$) at different corotation radii. From top to bottom: models  LRcor1.8, LRcor3, and
LRcor5.  \textit{Left panels:} 3D view at one of the density levels ($\rho$);
\textit{Middle panels:} bending waves ($z_w-$distribution); \textit{Right panels:}
density waves ($\sigma-$distribution).} \label{warp-160-all-9}
\end{figure*}

\subsubsection{Amplitude of the warp}
\label{sec:amplitude of warp}

It is important to know the amplitude of the warp.
     A large-amplitude warp can  block or
diminish the light from a star.
   In fact, this
mechanism of light-blocking by a warped inner disc has been proposed as an
explanations for dips in the light-curves observed in some T Tauri stars
\citep{bouvier03,bouvier07b,alencar10}.
    Clearly, the warp amplitude needs to be
sufficiently large to produce the observed dips.

     Figures \ref{warp-d1_5-c1_8-15} and \ref{warp-d0_5-c1_5-15} (middle panels)
show the amplitude of the warp, $z_w$ (the position of the center of mass of the disc
relative to the equatorial plane).
     Warp amplitude is the largest, reaching
$z_w\approx 0.2$, in model FW$\mu$1.5.
       Taking into account
the distance of this maximum to the star, $r\approx 2$, we obtain the ratio:
$z_w/r\approx 0.1$.
      In the model with a smaller magnetosphere,
FW$\mu$0.5, we find $z_w\approx 0.1-0.15$, and the ratio $z_w/r\approx 0.05$. These
numbers are in agreement with the theoretical analysis of warped discs by
\citet{terquem00}, who concluded that the amplitude of the warp is expected to be
$5-10\%$.
   However, they
considered the case of a very thin disc.

 In a realistic situation, the disc has a finite thickness, and the height of the warp
$h_w$ that can obscure light from the star can be larger than the deviation of the
center of mass from the equatorial plane, $z_w$. The density in the disc decreases
with height, and the height of the warp also depends on the chosen density level,
$h_w=h_w(\rho)$.

     To determine the height of the warp from our simulations, we chose
model FW$\mu$1.5, which has a larger magnetosphere.
 Fig. \ref{height-warp} (bottom panel) shows a view of the magnetosphere and the inner
warp, while the top panel shows a slice of the density distribution in the $xz$ plane,
which crosses the warp (note that the slice does not go through the maximum amplitude
of the warp, which is at about $30^\circ-40^\circ$ clockwise from the $xz-$plane).

    The top
panel shows that the height of the warp depends on the chosen density level: the lower
the density level, the higher the warp.
     In the figures, we chose a set of density levels.
    One can see that at the highest density levels (red color), the warp amplitude is not very
large, while at the lowest shown density level (dark blue) the warp rises up to $30\%$
of the distance to the star ($h_w/r=0.3$) and can be significant in blocking the light
at different inclination angles of observer.
      The warp amplitude is even larger at even lower density levels.
However, for an analysis of light obscuration by the warp, it is important to know the
opacity at different density levels in each particular case.
    We  note that the stellar light can also be obscured by the funnel streams, which are
located closer to the star than the warp, and can provide an even larger efficient
height of the obscuring material, $h_w/r>0.3$.

The relative amplitude $z_w/h_w$ and the height of the warp, $h_w/r$, depend on the
force which magnetosphere applies to the inner disc. It is expected that both values
should increase with the size of the magnetosphere, that is with $\tilde\mu$.
Comparison of amplitudes done in previous sections, shows that the ratio $z_w/r$ is
indeed twice as large in case of larger magnetosphere, $\tilde\mu=1.5$, than in small
magnetosphere, ($\tilde\mu=0.5$).

It may also depend on the temperature of the inner disc and its thickness.
       Fig. \ref{height-warp} shows that the accretion disc is
 thin, compared with the height of the warp and hence the magnetic force is the
 main one which determines the height of the warp in this model.

  We observed from our simulations that the ratios $z_w/r$ and
  $h_w/r$ are larger in model  FW$\mu$1.5 (with a larger magnetosphere, $r_m\approx 4.7 R_*$),
  compared with model FW$\mu$0.5 (where the magnetospheric radius is smaller, $r_m\approx 3.3 R_*$).
This is because a larger magnetosphere applies stronger magnetic force to the disc
(which has the same properties in these two models). We suggest, that an even stronger
warp can be expected in the cases of an even larger magnetosphere. It is expected that
in many CTTSs, the disc can be truncated at larger distances (e.g., $r_m\approx (7-8)
R_*$  in AA Tau star, \citealt{donati11}). Special investigations are needed to
understand
 whether the amplitude and height of the warp will increase or decrease with a further
 increase of $r_m$.

\begin{figure*}
\centering
\includegraphics[width=11.0cm]{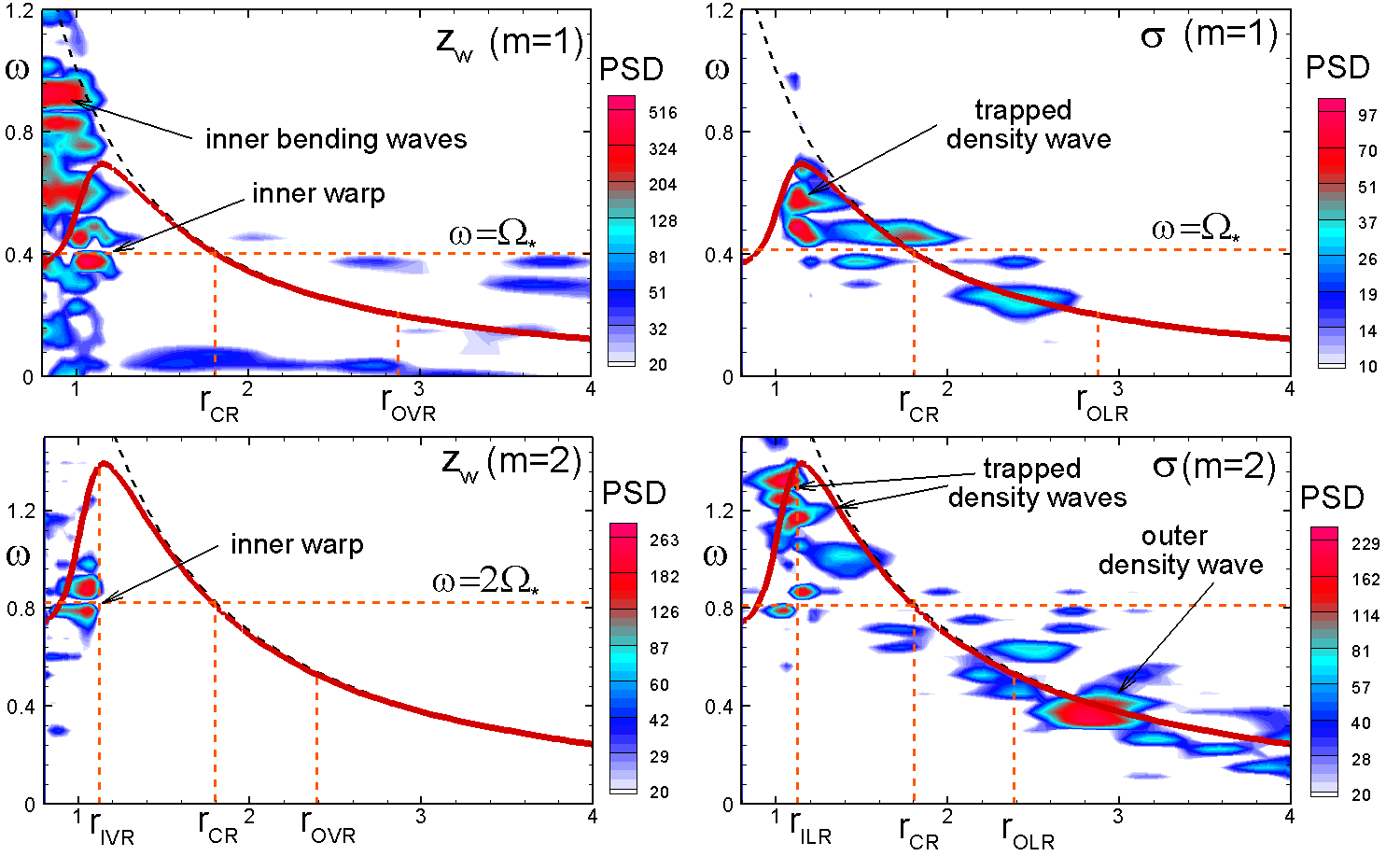}
\caption{PSD of oscillations (color background) for waves in model LRcor1.8 (large
region, $r_{cr}=1.8$). The solid red lines show the angular velocity distribution in
the disc} \label{psd-c1_8-160-r4-4}
\end{figure*}

\begin{figure*}
\centering
\includegraphics[width=11.0cm]{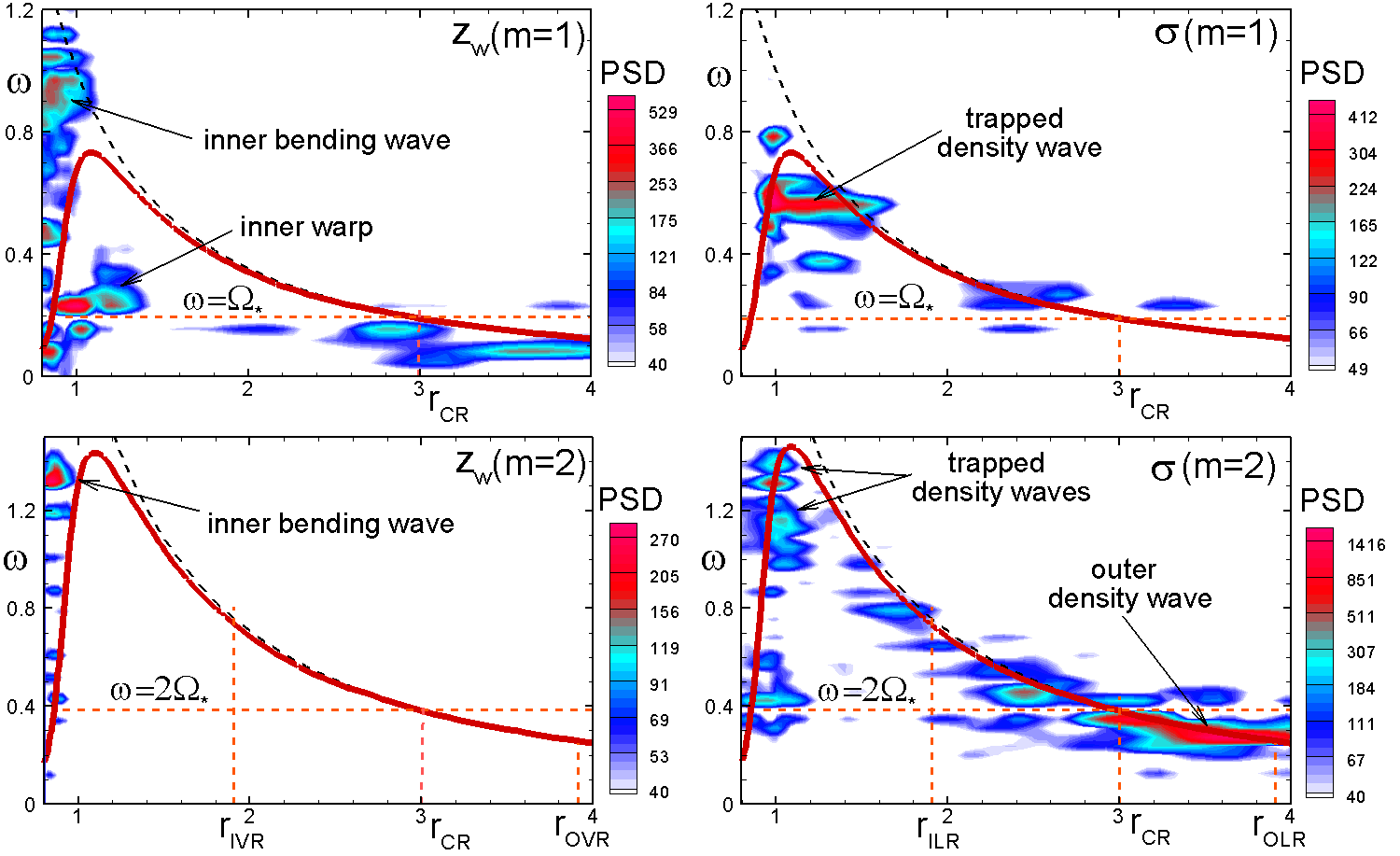}
\caption{Same as in Fig. \ref{psd-c1_8-160-r4-4}, but for model LRcor3 (large region,
$r_{cr}=3$).} \label{psd-c3-160-r4-4}
\end{figure*}

\begin{figure*}
\centering
\includegraphics[width=11.0cm]{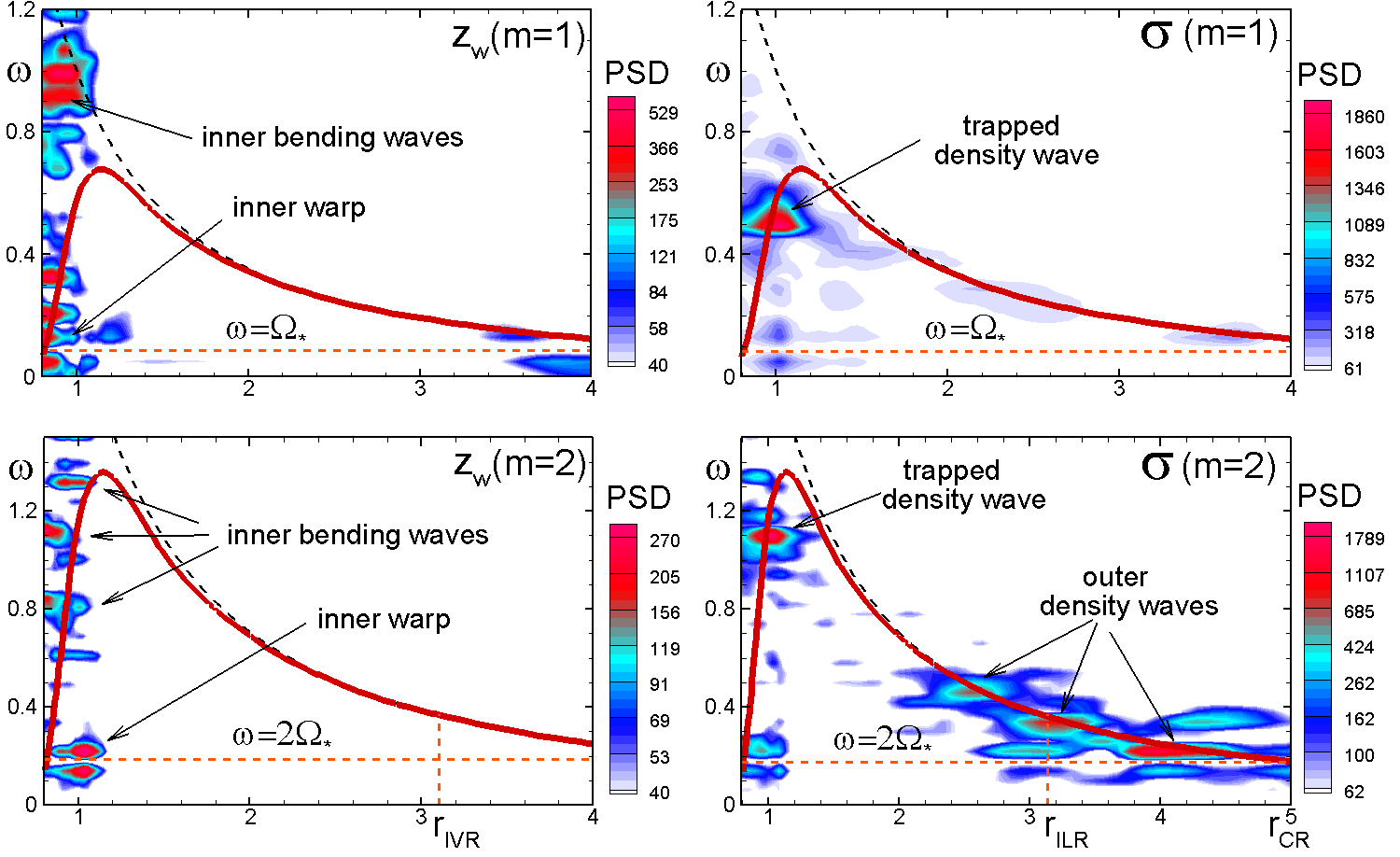}
\caption{Same as in Fig. \ref{psd-c1_8-160-r4-4}, but for model LRcor5 (large region,
$r_{cr}=5$).} \label{psd-c5-160-r4-4}
\end{figure*}

\subsubsection{Warps in the cases of different tilts of the dipole, $\theta$}
\label{sec:tilts}

Here, we investigate the dependence of the warp amplitude $z_w$ on the misalignment
angle of the dipole moment relative to the star's spin axis, $\theta$.
         For this reason, we performed a series of simulations at
different  $\theta$ values, from $\theta=5^\circ$ up to $\theta=90^\circ$ (models
FW$\theta$5, FW$\theta$15, FW$\theta$45, FW$\theta$60, and FW$\theta$90). We take as a
base model FW$\mu$0.5, where  $\theta=30^\circ$.

 Fig. \ref{warp-theta-5-15-30}  shows the
results for cases of relatively small tilts of the dipole: $\theta=5^\circ, 15^\circ$,
and $30^\circ$.
     The top panels show slices of density
distribution in the $xz-$plane and sample field lines. One can see that the inner disc
is thicker in models with smaller $\theta$. The densest part of the disc (red color)
shows bending waves. The middle panels show a 3D view of the bending waves at one
density level. One can see that waves are excited in all three models. The bottom
panels show the amplitude of the warp, $z_w$. A warp forms in all three cases, with
the maximum at the vertical resonance (outer dashed circle). From these plots, we find
the values $z_w/r\approx 0.025$, $0.045$, and $0.05$ in models with $\theta=5^\circ,
15^\circ$, and $30^\circ$, respectively.

The height of the warp, $h_w$ is larger than $z_w$. Top panels of Fig.
\ref{warp-theta-5-15-30} show that at small values of $\theta$ ($5^\circ$ and
$15^\circ$), the magnetosphere is an obstacle for the disc matter, and matter
accumulates at the disc-magnetosphere boundary, thus making the disc thicker. Such an
inner disc can absorb stellar light, thus providing $h_w/r\approx 0.3$. At
$\theta=30^\circ$, the amplitude of the warp $z_w$ is larger than in the other two
cases. However, the height of the inner disc is smaller.

The bottom panels  of Fig. \ref{warp-theta-5-15-30} show that in cases where
$\theta=5^\circ$ and $15^\circ$, some matter accretes to the star through a
Rayleigh-Taylor instability (e.g., \citealt{roma08}; \citealt{kulk08}). However, this
does not prevent excitation of bending waves, because these phenomena occur at
different radii: unstable accretion occurs at the inner regions of the disc, while
bending waves are excited at somewhat larger distances by the external parts of the
magnetosphere.

Fig. \ref{warp-theta-45-60-90} shows the excitation of bending waves around stars with
larger misalignment angles: $\theta=45^\circ, 60^\circ$ and $90^\circ$. Top two panels
show that bending waves are excited in all of these cases. However, the bottom panel
shows that a high-amplitude warp forms only when $\theta=45^\circ$ and $60^\circ$
(with ratios  $z_w/r =0.04$ and  $z_w/r =0.03$, respectively), while
 in the case of $\theta=90^\circ$, where the magnetosphere
 cannot provide strong perturbations in the $z-$direction, the ratio
 $z_w/r$ is about $10-20$ times smaller than that in the other two cases.

\subsubsection{Influence of viscosity on the warps} \label{sec:viscosity}

In our code an $\alpha-$type viscosity is incorporated into the disc. Viscosity is
thus regulated by the parameter $\alpha$ (see \citealt{shakura73}).
   All of the above-mentioned simulations
were performed for a relatively small viscosity, $\alpha=0.02$. Here, we show the
results of our simulations for the base model FW$\mu$0.5 but with different
viscosities: $\alpha=0, 0.04, 0.06$ and $0.08$.

Fig. \ref{warp-visc-4} shows bending waves for different   $\alpha$ values. One can
see that in all of the cases a warp forms, and is located at approximately the same
place, with a maximum amplitude between corotation and vertical resonances, or at the
vertical resonance\footnote{Here, for larger values of $\alpha$, we show the plot for
a later moment in time, $t$, because in this case the accretion rate is larger, the
disc has higher density, and bending waves are excited more slowly, compared with the
case of a smaller $\alpha$}.

    The possible influence of viscosity on the evolution of warped discs has been investigated
earlier (\citealt{papaloizou83,papaloizou95,terquem00,ogil06}). \citet{papaloizou95}
considered the formation of warps in viscous $\alpha-$type discs, which have a
vertical scale height $h(r)$, and concluded that to a first approximation, the warp
satisfies a wave-type equation if $\alpha \lesssim h/r$. In the opposite case, if
$\alpha \gtrsim h/r$, the warp satisfies a diffusion-type equation
\citep{papaloizou83}.

     In our simulations, the ratio $h/r\approx 0.1$.  Therefore, we expect non-diffusive,
     wave-like behavior for $\alpha$ values smaller than $\alpha=0.1$.
Hence, our range of parameters, $\alpha\le 0.08$, satisfies this
     condition. Simulations show that amplitudes of warps and bending waves
     are similar in all cases. This is somewhat unexpected result,
     because according to the theory, even in the wave regime, the
      viscosity should damp the waves (e.g., \citealt{ogil99,terquem00}).
       We do not see such damping. We suggest that
this issue requires further study and possibly a larger set of numerical simulations.

\subsection{Waves in cases where the magnetosphere rotates slower than the inner disc, $r_m < r_{cr}$}
\label{sec:slow warp section}

In the above section we investigated cases where the magnetosphere and the inner disc
rotate approximately with the same angular velocity, which corresponds to the
rotational equilibrium state (e.g., \citealt{long05}). We observed that the main
feature is a warp corotating with the star. However, the situation is different when
the magnetosphere rotates much slower than the inner disc, that is, when the
corotation radius $r_{cr}$ is much larger than the magnetospheric radius, $r_m$. Such
a situation may arise during periods of enhanced  mass accretion,  when the disc
compresses the magnetosphere and the magnetospheric radius $r_m$ decreases.
   Alternatively, the dipole component of the magnetic field of a young
star may decrease due to internal dynamo processes (e.g., \citealt{donati11}). As a
result, the magnetospheric radius $r_m$ will decrease, providing the condition $r_m <
r_{cr}$.

   If the magnetosphere rotates slower than the inner disc, then
it applies force to the inner disc, but with relatively low frequency of the star's
rotation, $\Omega_*$. This will excite bending and density waves with the frequency of
the star.
    On the other hand, the matter of the inner disc has a higher angular velocity.
It interacts with the slowly-rotating magnetosphere, which serves as a
non-axisymmetric obstacle for the disc matter flow. It exerts a non-axisymmetric force
on the inner parts of the disc, thus exciting higher-frequency waves (both density and
bending waves).
   Thus, a complex pattern of waves is expected in this case. In addition, we
observed that a slowly-rotating star excites \textit{global bending oscillations of
the whole disc} if the size of the simulation region (that is, the size of the disc)
is not very large. In the case of a large simulation region, these waves are also
excited, but their frequency is very low and they are excited slowly compared with the
overall time of simulations. This is why we initially consider the case of a
relatively large simulation region, where these low-frequency waves can be neglected
(see Sec. \ref{sec:slow-large-region}). Next, we investigate the cases of smaller
simulation regions, where global disc oscillations develop and couple with other
frequency modes (see Sec. \ref{sec:slow-small-region}).

\subsubsection{Investigation of waves in a large disc} \label{sec:slow-large-region}

Here, we investigate the waves excited in a relatively large simulation region, with
radius $R_{max}=20.2\approx 57.7 R_*$. This region is 1.7 times larger than that used
in Sec. \ref{sec:fast warp}. We observe that the very low-frequency global mode of
disc oscillations develops slowly, and we are able to investigate higher-frequency
oscillations at the ``background" of this slowly-growing mode.

We use a model with parameters corresponding to our base model FW$\mu$0.5, but take a
larger simulation region and a larger corotation radius: $r_{cr}=1.8, 3$ and $5$
(models LRcor1.8, LRcor3, and LRcor5).
    The magnetospheric radius $r_m\approx 1.15$ is located at the
same place as in model FW$\mu$0.5.
   Hence, the new models correspond to stars with slower rotation.
    The simulations show that matter in the rapidly-rotating inner disc interacts with
the slowly-rotating tilted magnetosphere of the star, so that different types of waves
are excited and propagate to larger distances.

Fig. \ref{warp-160-all-9} (top row) shows the results of our simulations for model
LRcor1.8, where the corotation radius is $r_{cr}=1.8$, and the star rotates relatively
rapidly compared with the other two cases.
    The left and middle panels show that a bending wave is excited at the
disc-magnetosphere boundary and propagates outward, forming
a spiral wave. The amplitude is enhanced at the outer vertical resonance, $r=r_{\rm
OVR}$ (see Table \ref{tab:resonances}), and this enhancement has a position similar to
that of the warp in the base case (FW$\mu$0.5), where the wave corotates with the
star. Here, however, instead of corotation, we observe \textit{a tendency} to
corotate, where a wave rotates with the frequency of the star during only a part of
the rotational phase, then it slows down, but soon after, another similar wave appears
and corotates with the star, and so on. The right panel shows surface density
distribution, $\sigma$. One can see that different waves form at different radii from
the magnetosphere, including a two-armed density wave, which forms at the outer
Lindblad resonance.

Next, we show the results for model LRcor3, where the corotation radius is larger,
$r_{cr}=3$. Fig. \ref{warp-160-all-9} (middle row)) shows that bending waves are less
ordered inside the corotation radius compared with model LRcor1.8.
   They look like parts of a
tightly wound spiral wave extending out to the corotation radius.
   At larger distances, a new
bending wave is formed, and it is more ordered and not as tightly wound.
    This observation is
in agreement with the theoretical prediction, where a tightly wound spiral wave is
expected at $r<r_{cr}$ (see the left two panels of Fig. \ref{Appfig:allwarps} in
Appendix \ref{sec:Appendix-allwaves-modes}), and a much less tightly wound spiral is
expected at $r>r_{cr}$ (see Fig. \ref{Appfig:allwarps}, right panels).
    The right-hand panel of Fig. \ref{warp-160-all-9} (middle row)  shows that different density waves
formed between the inner radius of the disc and the OLR radius.

Next, we show the results for model LRcor5, where the corotation radius is even
larger, $r_{cr}=5$. Fig. \ref{warp-160-all-9} (bottom row) shows that bending waves
are not ordered and have a rippled structure, as in the $r_{cr}=3$ case. The middle
panel shows that bending waves are not ordered, but one can track a tightly wound
spiral wave inside the corotation radius (in agreement with Fig. \ref{Appfig:allwarps}
of Appendix \ref{sec:Appendix-allwaves-modes}). A more ordered, but less tightly wound
spiral wave starts at $r\approx r_{cr}$ and becomes yet more ordered at the OVR. The
right-hand panel shows density waves. One can see that different density waves
propagate at $r<r_{ILR}$. There are waves at larger distances, but they are not seen
at the chosen density levels.

The PSD shows interesting sets of bending and density waves in these three models (see
Figures \ref{psd-c1_8-160-r4-4}, \ref{psd-c3-160-r4-4} and \ref{psd-c5-160-r4-4}).
   The PSD confirms that there is no large-scale warp at the stellar
frequency, which is different from models FW$\mu$0.5 and FW$\mu$1.5 (see Sec.
\ref{sec:fast warp}).
   Instead, only a small-scale bending wave is excited at
the stellar frequency or at twice this value (see left panels of the figures for $m=1$
and $2$).
   This wave is only present at the inner region of the disc, close to the region
where matter  lifted to the funnel flow, and may therefore reflect this lifting. For
the three models considered, the angular frequency of the wave is low. For example, in
the case of a one-armed wave ($m=1$), $\omega\approx 0.41, 0.19$ and $0.09$ for models
LRcor1.8, LRcor3 and LRcor5, respectively.

There are also high-frequency bending waves associated with interaction of the
rapidly-rotating matter of the inner disc with the non-axisymmetric magnetosphere. The
angular frequency of these waves, $\omega\approx 0.8-1$, approximately corresponds to
Keplerian velocity of the inner disc. Also, there are bending waves of intermediate
frequency. All of these waves are located at the inner edge of the disc, $0.8<r<1.2$
where matter has already been lifted, or has started lifting to the funnel flow.

Lower-frequency bending waves excited by the slowly-rotating star propagate from the
disc-magnetosphere boundary to large distances, and we often see an enhanced PSD at
the stellar frequency, and at different distances from the star.  The PSD can be
further enhanced at the corotation radius, as we can see in Fig. \ref{psd-c3-160-r4-4}
(top left panel). The stellar-frequency mode is also seen in other models, but at
lower values of PSD.

The density waves  show a variety of frequencies (see right-hand panels of figures
\ref{psd-c1_8-160-r4-4}, \ref{psd-c3-160-r4-4} and \ref{psd-c5-160-r4-4}). At low
frequencies, we observe enhanced PSD at radii $\sim 3-4$, which often correlate with a
position of different resonance (e.g., in model LRcor1.8 with OLR, in model LRcor3
with CR and OLR, in model LRcor5 - with ILR).
  Plots of the surface density
distribution $\sigma$ for these models (see Fig. \ref{warp-160-all-9}) show that in
all these models, the density waves have higher amplitudes at the distances of
$r\lesssim 4$, where the influence of the rotating magnetosphere is stronger.


\begin{figure*}
\centering
\includegraphics[width=16.0cm]{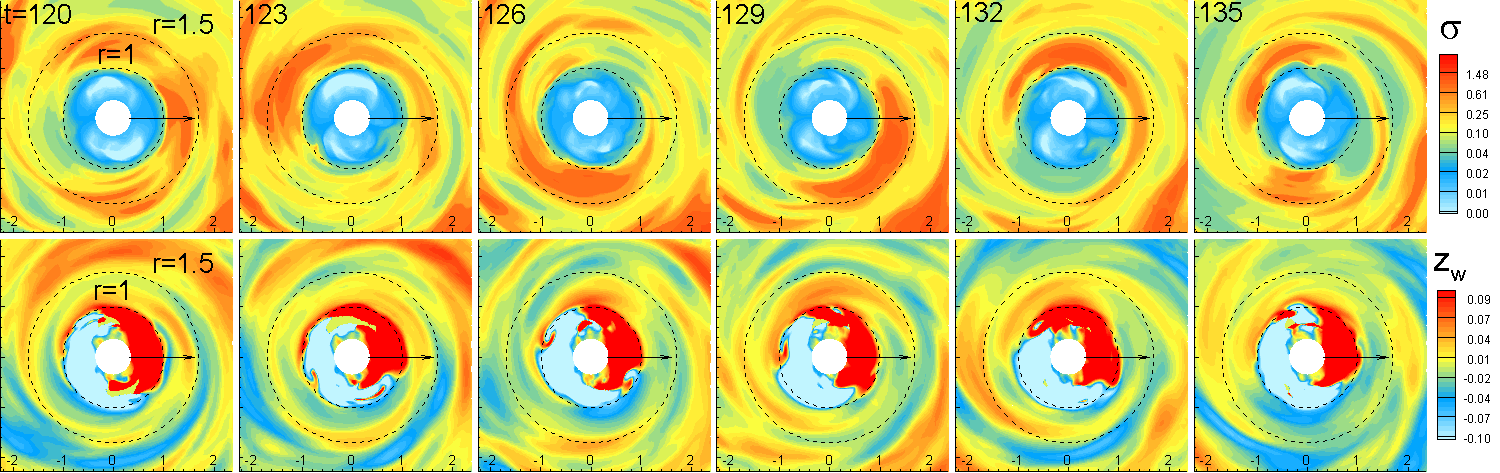}
\caption{Evolution of density (top panels) and bending (bottom panels) waves at the
inner disc in model LRcor3  (large region, $r_{cr}=3$). Dashed circles show radii
$r=1$ and $r=1.5$ (for scale).} \label{sigma-warp-c3-12}
\end{figure*}

\begin{figure*}
\centering
\includegraphics[width=16.0cm]{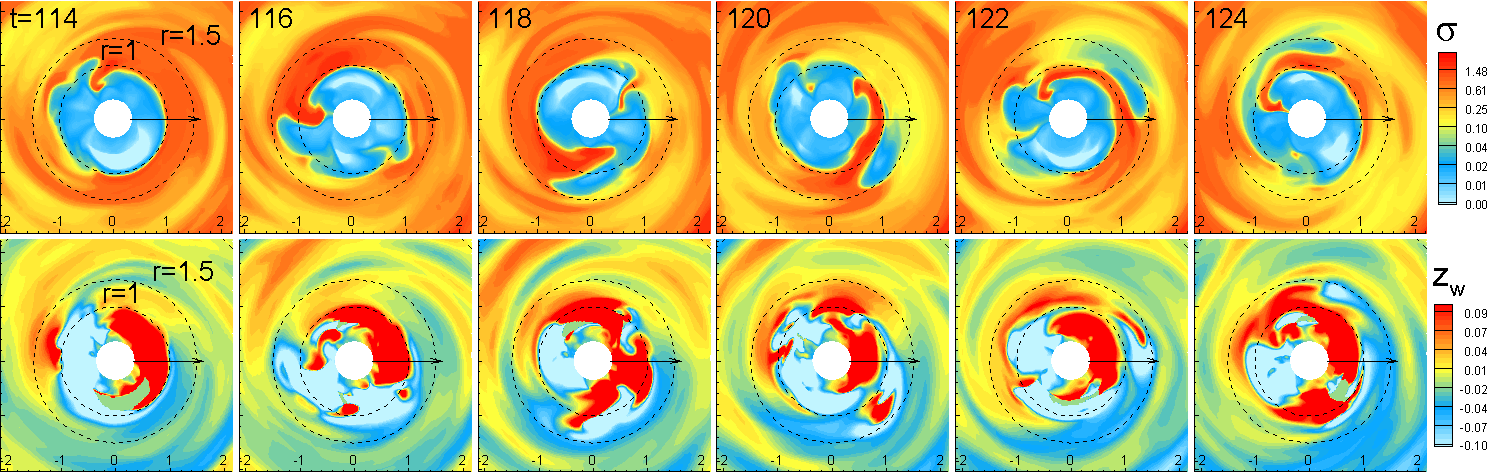}
\caption{Same as in Fig. \ref{sigma-warp-c3-12}, but for model LRcor5 (large region,
$r_{cr}=5$).} \label{sigma-warp-c5-12}
\end{figure*}

\subsubsection{Waves in the inner disc: trapped density waves}
\label{sec:trapped waves}

At high frequencies, the PSD of density waves is often enhanced at radii corresponding
to the peak in the angular velocity distribution, and the frequencies of these waves
are usually somewhat lower than the peak frequency  (see Fig. \ref{psd-c1_8-160-r4-4},
\ref{psd-c3-160-r4-4} and \ref{psd-c5-160-r4-4}, where the solid red lines show the
angular velocity distribution in the disc).
    We suggest that these are \textit{trapped} density waves,
which are expected in cases where the angular velocity distribution in the disc has a
maximum \citep{love07,love09}.
     In our models, the maximum in the angular velocity as a function of $r$ occurs
     because the star rotates slower than the inner disc.
    The region where the angular velocity decreases gives rise to the possibility of
radially trapped unstable Rossby waves with azimuthal mode numbers $m\geq 1$ and
radial width $\Delta r/r \ll1$ (\citealt{love09}).
 The theory of trapped waves is summarized
in Appendix \ref{sec:Appendix-trapped modes}. It is important that the Rossby waves
have density variations as well as temperature variations, and hence they appear in
our simulations and  analysis as density waves.
      Here, we analyze trapped waves in one of our models
(LRcor3), where these waves are clearly observed.
     Fig. \ref{psd-c3-160-r4-4} shows a
high-frequency ($\omega\approx 0.57$), one-armed ($m=1$) wave, which is located below
the maximum in $\omega$ at radii $1<r<1.5$. To check this wave we also plot the
surface density distribution at the inner disc for different times and clearly see
that there is a one-armed wave located approximately between radii 1 and 1.5 (see Fig.
\ref{sigma-warp-c3-12}).
    We believe that this is a trapped density wave.
According to the theory (\citealt{love09}), trapped waves have a tendency to be within
the maximum in the $\omega$ distribution. We can see that in  model LRcor3, the
maximum of the PSD for $m=1$ is located in the left part of the $\omega-$curve, while
for model LRcor5, the trapped wave is located below the maximum
 of $\omega$ for both the $m=1$ and $m=2$ modes.

The bottom panels of Fig. \ref{sigma-warp-c3-12} show bending waves for the same
moments in time. A small inner warp forms at $1<r<1.5$, but rotates  with the angular
velocity of the star. This small warp is also seen in Fig. \ref{psd-c3-160-r4-4} (top
left panel).
     The PSD also shows the presence of a high-frequency,
($\omega\approx 1$), bending wave at $r < 1$ (top left panel). This feature probably
corresponds to a small funnel or tongue, which forms due to the \textit{magnetic
Rayleigh-Taylor (RT) instability} at the disc-magnetosphere boundary.

\begin{figure*}
\centering
\includegraphics[width=16cm]{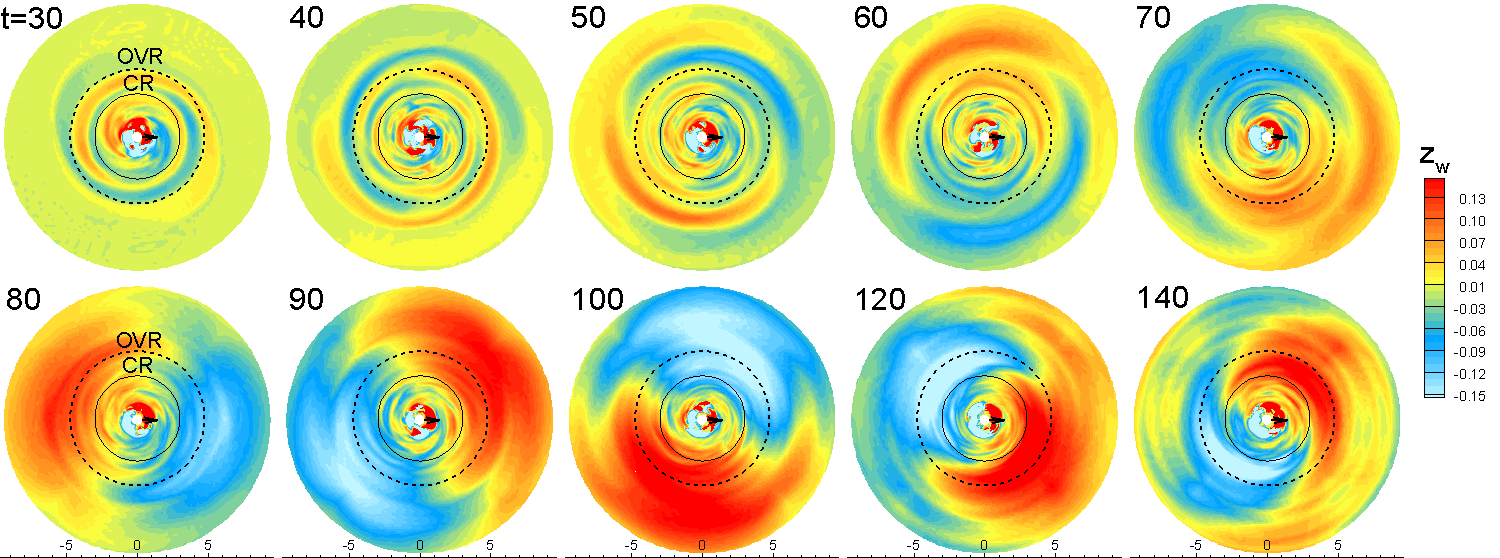}
\caption{The sequence of panels shows how a bending wave near the star excites a
global oscillation of the whole disc. Simulations are performed in a relatively small
region, $R_{out}=9.4=26.9 R_*$, for demonstration (model SRcor3).}
\label{warp-c3-130-10}
\end{figure*}

\subsubsection{Accretion through instabilities: density and bending waves}
\label{sec:instabilities}

At the disc-magnetosphere boundary, matter  can penetrate through the magnetosphere
due to the magnetic Rayleigh-Taylor (RT) instability (e.g.,
\citealt{arons76,spruit90,kaisig92}). A necessary condition for the instability is
that the gravitational acceleration towards the star be larger than the centrifugal
acceleration.
   Additional factors are also important, such as the
gradient of the angular velocity in the inner disc and the gradient of the surface
density (e.g., \citealt{spruit95}).
    Global numerical simulations of accretion onto a star with
a tilted dipole magnetic field show that accretion may be in the stable or unstable
regimes (see \citealt{roma08,kulk08,kulk09,bac10}).
    In the stable regime, matter accretes
in ordered funnel streams, which corotate with the star.
   In the unstable regime, matter
accretes through unstable temporary funnels or tongues, which carry the angular
momentum of the inner disc, and can rotate faster than the inner disc matter.
     The simulations show
that the criterion
 for the onset of the RT instability is close to that derived by
\citet{spruit95}. In addition, the boundary between stable and unstable regimes
depends on the tilt of the magnetosphere and on the simulation grid \footnote{Recent
simulations performed with a finer grid (as in this paper) show that the RT
instability occurs more readily compared with the earlier simulations performed with a
coarser grid (e.g., \citealt{roma08}).}. At larger tilt angles, more matter accretes
in stable funnel streams.

 In our models with $r_{cr}=3$ and $5$, a star rotates slowly relative to the inner disc,
  and therefore the instability is favored. On the other hand, the tilt of
 the dipole is relatively large ($\theta=30^\circ$), and this favors accretion through funnel streams.
     This is why
in the case of $r_{cr}=3$, where the star rotates relatively slowly, we see only a
weak instability.
    To demonstrate accretion through instabilities, we take an even slower rotating
star, where $r_{cr}=5$ (model LRcor5) and accretion through the RT instability is very
clear.
    The top panels of Fig. \ref{sigma-warp-c5-12} show that the instability leads to
the formation of one main ``tongue", which rotates with the angular frequency
$\omega\approx 0.5$. This unstable tongue is associated with the main one-armed
trapped density wave that forms in the inner disc. The frequency of this trapped wave
is clearly seen in Fig. \ref{psd-c5-160-r4-4} (top right panel).  The PSD also shows
the presence of a second harmonic ($m=2$) with $\omega\approx 1.1$ at radius $r\approx
1$ (see bottom right panel of Fig. \ref{psd-c5-160-r4-4}). This wave is barely seen in
the surface density plots in Fig. \ref{sigma-warp-c5-12}, but is clearly seen in the
PSD.

The bottom panels of Fig. \ref{sigma-warp-c5-12} show that the instability is also
seen in the bending waves that form at the inner edge of the disc. The two-armed
bending wave (bottom left panel of Fig. \ref{psd-c5-160-r4-4}) also has a mode
$\omega\approx 1.15$ at $r<1$, which is close to the mode $\omega=1.1$ of the $m=2$
trapped density wave. Here, we have an example of how a trapped density wave generates
unstable tongues at smaller radii. Another possible explanation is that both waves
indicate accretion through instabilities.

Fig. \ref{psd-c5-160-r4-4} (top left panel) shows the presence of a high-frequency
($\omega\approx 0.9-1.1$), one-armed ($m=1$) bending wave at $r<1$. A similar mode is
observed in two other models (LRcor1.8, LRcor3). This mode is probably connected with
the rotation of unstable funnel streams at the inner edge of the disc. The frequency
of the mode is higher than that of the disc (solid red line). We suggest that this
mode may reflect the rotation of temporary unstable funnel streams, which are lifted
above the magnetosphere, and therefore do not experience magnetic braking (as matter
in the disc does), and hence can rotate with a nearly Keplerian velocity. We call
these waves \textit{inner bending waves}.

\subsection{Waves in a smaller simulation region: global disc oscillations and slowly-rotating warps}
\label{sec:slow-small-region}

In sections \ref{sec:slow-large-region}-\ref{sec:instabilities}, we considered the
case of a relatively large region, where free bending oscillations in the disc develop
slowly. In this section, we consider a smaller simulation region, in which we observe
that these oscillations develop more rapidly.

 First, we demonstrate the
excitation of free bending oscillations in the disc of a very small size, $R_{out}=
9.4=26.9 R_*$ (model SRcor3).
    The simulations show that
initially, waves are excited at the disc-magnetosphere boundary (see top panels of
Fig. \ref{warp-c3-130-10}).
   Later, at $t\approx 70-100$, global bending
oscillations of the disc  develop. However, at $t>100$, the maximum of the bending
wave amplitude moves to the region between the OVR and the CR resonances (see Fig.
\ref{warp-c3-130-10}). This feature is similar to the warps  in cases of
rapidly-rotating stars (see Sec. \ref{sec:fast warp}). However, this new warp rotates
\textit{slower} than the star.

Next, we investigate the waves in the intermediate-sized simulation region, $R_{out}=
12.1=34.5 R_*$, where free bending oscillations in the disc develop sufficiently fast
in particular in cases of slowly-rotating stars).  We consider two models with
corotation radii $r_{cr}=1.8$ and $3$ (models SWcor1.8 and SWcor3).

\begin{figure*}
\centering
\includegraphics[width=16.0cm]{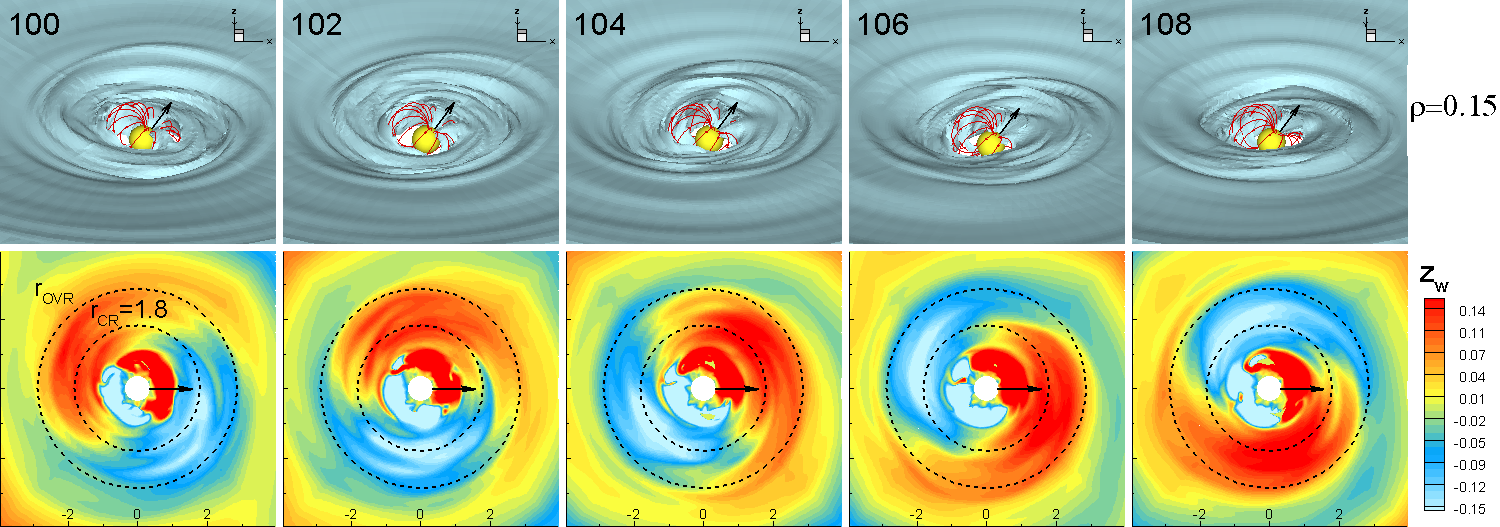}
\caption{A slowly-rotating warp observed in model SWcor1.8. \textit{Top row:} 3D views
of the warp at different moments in time; \textit{Bottom row:} center of mass of the
disc $z_w$, which demonstrates a slowly-rotating warp.} \label{warp-c1_8-140-10}
\end{figure*}

\begin{figure*}
\centering
\includegraphics[width=17.0cm]{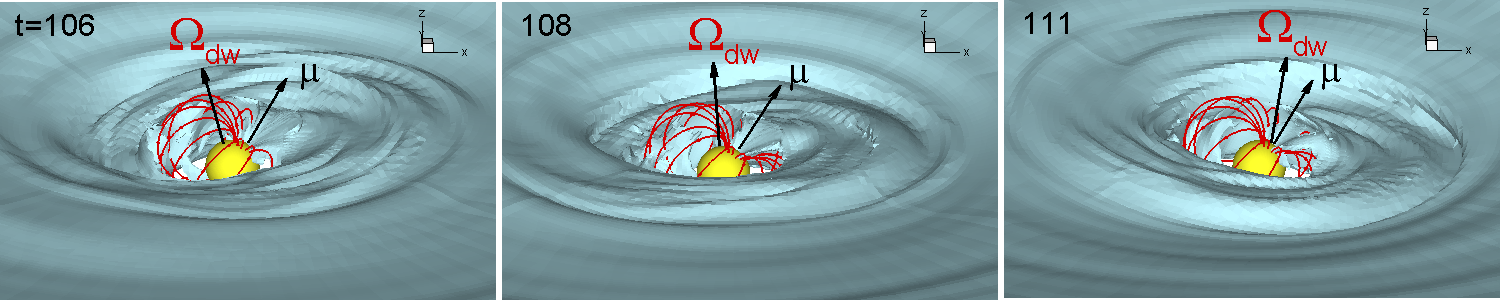}
\caption{An example of a warp which rotates slower than the star in model SWcor1.8
(simulation region $R_{out}=12.1=34.5R_*$). One of the density levels is shown. Vector
$\Omega_{dw}$ shows the direction perpendicular to the inner disc warp.}
\label{warp-c1.8-3d-3}
\end{figure*}

In model SWcor1.8, we observed evolution of bending waves similar to that in model
SRcor3 (see Fig. \ref{warp-c3-130-10}). As a result of this evolution, a warp forms at
radii $r_{\rm CR}<r<r_{\rm OVR}$.
   This warp rotates slower than
the star (see Fig. \ref{warp-c1_8-140-10}). A closer view of the warped disc (see Fig.
\ref{warp-c1.8-3d-3}) shows that a significant part of the inner disc is tilted and
involved in this slow rotation.

\begin{figure*}
\centering
\includegraphics[width=16.0cm]{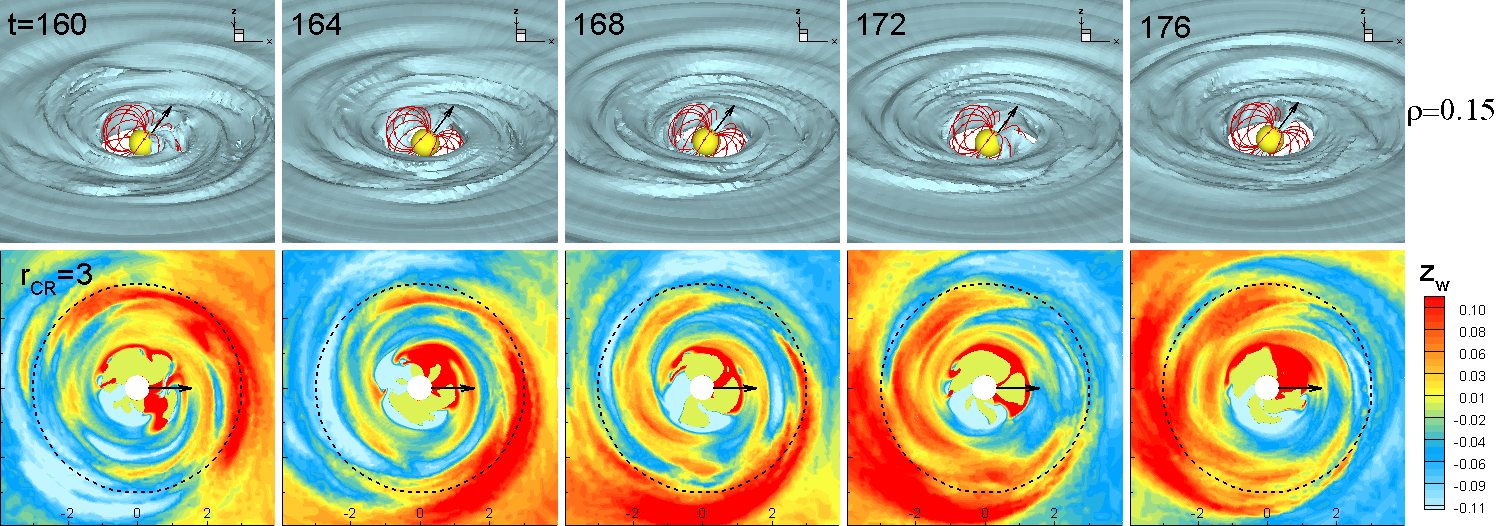}
\caption{Same as in Fig. \ref{warp-c1_8-140-10}, but for a slower rotating star,
$r_{cr}=3$ (model SWcor3).} \label{warp-c3-140-10}
\end{figure*}

In the case of an even more slowly-rotating star, $r_{cr}=3$ (model SWcor3), a very
slow warp forms at $r>r_{\rm CR}$.  Fig. \ref{warp-c3-140-10} shows the rotation of
this warp about the star.

\begin{figure*}
\centering
\includegraphics[width=12.0cm]{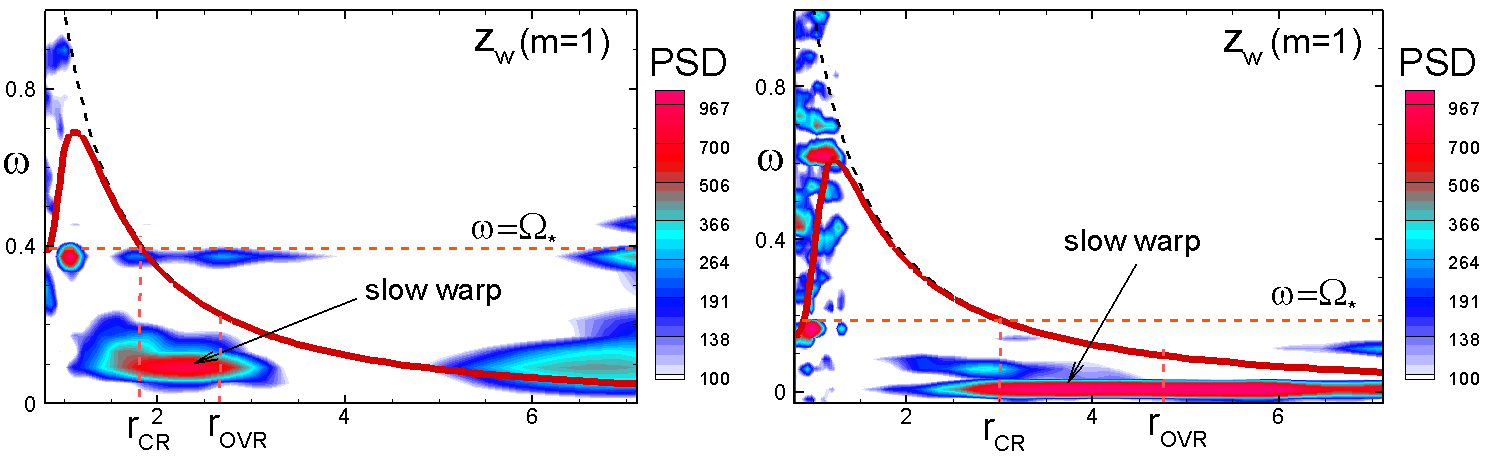}
\caption{PSD demonstrates low-frequency mode associated with a slow, one-armed ($m=1$)
bending wave in models SWcor1.8 (left panel) and SWcor3 (right panel).}
\label{psd-c1_8-c3-140-2}
\end{figure*}

Fig. \ref{psd-c1_8-c3-140-2} shows the PSD for these two models. One can see that in
the case where  $r_{cr}=1.8$ case (left panel), the warp rotates with an angular
frequency $\omega\approx 0.1$, which is about four times smaller than frequency of the
star.
   The PSD also shows that the main part of the warp is located between the corotation and the
   vertical resonances.
   In the case of a slower rotating star ($r_{cr}=3$, right panel), the warp rotates very slowly,
    with an angular frequency
    $\omega=0.01-0.02$, though the accuracy of our model is not very high in cases of
    very low frequencies (because the number of rotations per simulation run is
small).
     One can see that in both cases, the angular
frequency of slow warps is much smaller than the local Keplerian frequency. In model
SWcor3, the angular frequency of the warp is comparable with the Keplerian frequency
at the edge of the simulation region, $\omega_K(r\approx 12)\approx 0.024$.
   However, this is probably just
a coincidence. In test simulations with even more slowly-rotating stars  ($r_{cr}=5$),
a warp forms, and its angular frequency is even lower: $\omega<0.01$.
   We did not establish
how this slow warp forms. Its frequency probably represents a coupling between the
frequency of the star and  the frequency of global oscillations in the disc.

\section{Applications and discussion}
\label{sec:applications-discussion}

The results of our simulations can be applied to various types of stars where the
magnetospheric radius is several times larger than the radius of the star
\footnote{Modeling of much larger magnetospheres with truncation radii $r_m>>10R_*$,
requires much longer simulation runs and is at present not practical. Special efforts
are required to investigate stars with large magnetospheres, such as X-ray pulsars.},
including, for example, classical T Tauri stars, accreting millisecond pulsars that
host spun-up neutron stars and some types of cataclysmic variables (such as Dwarf
Novae). Possible applications are briefly discussed in sections below.

\subsection{Obscuration of light by warped discs in classical T Tauri stars (CTTS) and drifting period }
\label{sec:application-ctts}

The light curves of classical T Tauri stars (CTTS) display photometric, spectroscopic
and polarimetric variations on timescales ranging from a few hours to several weeks
\citep{herbst02}.
   \citet{bouvier99} studied the variability of AA Tau in great detail and
concluded that the photometric variability with a period of $8.2$ days which is
comparable to the expected rotation period of the star is due to the occultation of
the star by a warp formed in the inner disc (the system is observed almost edge-on).
      They
proposed that this warp is produced by the interaction of the disc with the stellar
magnetic dipole tilted with respect to the disc rotational axis.
    Later, more detailed
analysis confirmed the hypothesis of obscuration by the disc
\citep{bouvier03,bouvier07b}. Doppler tomography observations of AA Tau show that the
dominant component of the field is a $2-3$ kG dipole field that is tilted at
$20^\circ$ relative to the rotational axis. At this field strength the magnetospheric
radius is close to the corotation radius: $r_m\approx r_{cr}$ \citep{donati10}.
   Our simulations show that in this situation
a large amplitude warp forms and rotates with the frequency of the star, and the AA
Tau warp may be similar to that described in model FW$\mu$1.5.
    An analysis of the warps in cases where the dipole has different tilts
    (see Sec. \ref{sec:tilts}) shows that an angle of $\theta\lesssim 20^\circ$ is sufficiently
large to generate a warp. It is possible that warps are common features in many CTTSs.
For example, \citet{alencar10} analyzed the photometric variability of CTTSs in the
young cluster NGC 2264 using data obtained by the \textit{CoRoT} satellite and
concluded that AA Tau-like light curves are  fairly common, and are present in at
least ~$30-40\%$ of young stars with inner dusty discs (see also
\citealt{carpenter01}).

Not only fast warps, but also other waves can be generated by the rotating
magnetosphere in CTTSs. However, it is difficult to observe these waves in CTTSs:
these stars are usually brighter than the inner disc and are strongly variable due to
coronal magnetic activity. In addition, observations usually cover not more than a few
periods of the star, and it is difficult to extract a possible frequency of the waves
from light-curves. One possibility is that the waves generated in the inner disc may
influence the formation of magnetospheric streams and their subsequent motion, thereby
leaving a trace on the surface of the star in the form of hot spots, which can move
faster or slower than the star
 \citep{roma04,bac10}. Observations show that many CTTSs do not have a precise period, but instead
a quasi-period, which varies
  with time around some value (e.g., \citealt{rucin08}). This
quasi-period may be connected with the formation of high-frequency waves, such as
trapped density waves. These waves represent regions of enhanced density and therefore
probable places where funnel streams may form. The position and frequency of a trapped
waves vary with accretion rate.  Thus, the angular velocity of hot spots on the
surface of the star and the associated quasi-period of the star will also vary.

In a number of situations accretion through the RT instability is possible, where
matter accretes in unstable, stochastically-forming tongues (e.g.,
\citealt{roma08,kulk08}). In this case, the light-curve becomes less ordered, and in a
strongly unstable case a CTTS may show the absence of a period. It is often the case
that the instability is not very strong, and both stochastic and periodic components
are expected in the frequency spectrum \citep{kulk09}.  In these cases, trapped
density waves can determine the position of unstable tongues. In the case of small
magnetospheres, this leads to the formation of two coherent tongues that rotate more
rapidly than the star \citep{roma09}. Hence, one of the frequencies can be connected
with trapped density waves. On the other hand, we often observe in the PSD even
higher- frequency oscillations (inner bending waves) that have a nearly Keplerian
frequency. Both of these waves may leave an imprint at the surface of CTTSs in the
form of moving hot spots. Our simulations show that the coherence of trapped waves is
probably higher than that of inner bending waves, and it is more likely that the
variable quasi-period of CTTSs is connected with the trapped waves.

\subsection{Application to Accreting Millisecond Pulsars}
\label{sec:application-MP}

Accreting millisecond pulsars (AMPs) show various types of quasi-periodic oscillations
(QPOs), ranging from very high frequencies ($\sim 1,300$ Hz) down to very low
frequencies ($\sim 0.1$Hz) (e.g. \citealt{vanderKlis06}). One of the prominent
features in the spectrum of AMPs is a pair of QPO peaks with an upper frequency
$\nu_u$ and a lower frequency $\nu_l$, which move in pairs. The distance between the
peaks $\nu_u-\nu_l$ usually corresponds to either the frequency of the star, $\nu_*$,
to half this value, $\nu_*/2$ (e.g., \citealt{vanderKlis00,vanderKlis06}), or, in some
AMPs there is no such correlation (e.g., \citealt{mendez98,bout06,bel07,mendez07}).
There are also QPOs with lower frequencies, including hecto-hertz QPOs with a
frequency of $\nu_h\approx \nu_u/4$, as well as low frequency QPOs with a frequency of
$\nu_{\rm LF}\approx \nu_u/15$.
 There are also very low-frequency oscillations, $\nu_{LL}\lesssim
1$Hz. Below we discuss possible connections between the waves observed in simulations
and some of these frequencies.


\subsubsection{High-frequency oscillations in the inner disc}
\label{sec:AMP-high frequency}

Our simulations show that there are two main types of high-frequency waves: (1)
 \textit{trapped density waves}, and (2) \textit{inner bending waves}. The frequency
of the inner bending waves approximately equal to the Keplerian velocity at the
disc-magnetosphere boundary. The trapped density wave is located inside the peak in
the disc angular velocity distribution, and hence it has lower frequency.  We suggest
that the trapped waves can be responsible for the lower QPO peak, $\nu_l$, and the
inner bending wave for the upper QPO peak, $\nu_u$.

Variation of the accretion rate leads to variation of the position of the
disc-magnetosphere boundary, and therefore the position of both peaks will vary with
accretion rate and they will move in pairs. We expect that the quality factor
$Q=\nu/\Delta \nu$  of both waves will increase with accretion rate, because when the
disc moves closer to the star, the maximum in the angular velocity distribution
becomes more pronounced, and the trapped waves become stronger and more coherent. At
the same time, accretion through instabilities becomes stronger and more ordered
(where one or two tongues become dominant and create a more ordered motion, see
\citealt{roma09}). This is in agreement with the observations of QPOs, where $Q$
increases with frequency for both upper and lower QPOs (e.g.,
\citealt{vanderKlis00,barret06}). The quality factor of lower QPOs starts decreasing
at very high frequencies, possibly due to relativistic effects (e.g.,
\citealt{barret06}).

Waves in the inner disc can influence the magnetospheric flow and can therefore leave
imprints on the frequencies of moving spots on the surface of the star. Moving spots
on the surface of the star were observed in global 3D simulations of stable
magnetospheric accretion \citep{roma03,roma04} and during accretion through
instabilities (e.g., \citealt{kulk08}). \citet{bac10} found that rotating spots
provide two QPO frequencies: one corresponds to rotating funnel streams and has a
lower frequency, while the other corresponds to rotating unstable tongues and has a
higher frequency (see also \citealt{kulk09}). We suggest that these two frequencies
represent an imprint of the trapped and inner bending waves, respectively.

We can calculate the difference in frequency between these two high-frequency waves in
the disc. For example, in model LRcor3, the frequency of the $m=1$ trapped wave is
$\omega_{trap}\approx 0.55$, while the frequency of the inner bending wave is
$\omega_{bend}\approx 0.95$ (see top panels in Fig. \ref{psd-c3-160-r4-4}). Then,
using Table \ref{tab:refval} for millisecond pulsars, we obtain the dimensional
frequencies of these waves, $\nu_l=247$Hz and $\nu_u=427$Hz, and the difference
between them, $\Delta\nu=180$Hz. Similarly, in model LRcor5 (see top panels of Fig.
\ref{psd-c5-160-r4-4}) we obtain $\omega_{trap}\approx 0.51$ ($\nu_l\approx 229$Hz)
for the trapped density waves, and two candidate frequencies for the inner bending
waves, $\omega_{ bend}\approx 0.91$ and $\omega_{bend}\approx 0.99$, which correspond
to $\nu_u=409$Hz and $\nu_u=445$Hz. The difference between the frequencies is $180$Hz
and $216$Hz, respectively. This difference is typical for many accreting millisecond
pulsars (see bottom panel of fig.1 of \citealt{mendez07}).

For case of moving spots, we investigated how frequencies vary with accretion rate and
found that the typical difference is $\Delta\nu\approx 200-250$Hz, which is in
agreement with the frequency difference for high-frequency waves. This strengthens our
hypothesis that the hot spots represent an imprint of the waves in the disc.


We should note that the frequency difference $\Delta\nu$ is approximately the same for
stars with $r_{cr}=3$ as for stars with $r_{cr}=5$, and therefore $\Delta\nu$ does not
depend on the frequency of the star. This model can explain those AMPs where
$\Delta\nu$ does not correlate with the frequency of the star (e.g.,
\citealt{mendez98,bout06,bel07,mendez07}).

It has been suggested earlier that the high-frequency QPOs may be connected with the
rotation of some blobs in the inner parts of the disc \citep{lamb85}. The position of
the inner disc can be determined, for example, by the magnetospheric radius (e.g.,
\citealt{roma09}, \citealt{bac10}), or by a sonic point \citep{mill98}. In our
simulations, instead of blobs we observe \textit{trapped density waves} or inner
bending waves. These waves are more ordered than blobs and can provide a much higher
quality factor.

\subsubsection{Frequency Modulation at the Vertical Resonance}

In many AMPs, the difference between the QPO peaks is equal to the frequency of the
star $\nu_*$ or to $\nu_*/2$ (e.g., \citealt{vanderKlis00}). Here, we discuss a
possible model that follows from our simulations and that satisfies to one or both
conditions.

A theoretical analysis of the disc waves shows that if waves are excited by the
vertical force oscillating with the frequency $\nu_f$, then the largest amplitude of
the waves is expected at the outer vertical resonance, where the frequency of the wave
is $\nu_{\rm OVR}= \nu_f/2$. \citet{lamb03} proposed a model where the upper
frequency, $\nu_u$, is connected to the rotation of blobs in the inner disc, while the
lower frequency $\nu_l$ is connected to the modulation of light at this resonance,
$\nu_l=\nu_u-\nu_f/2$. In the case of a tilted rotating dipole magnetosphere, the
forcing frequency can be either one or two times the frequency of the star, and
therefore the difference in frequencies $\nu_u-\nu_l$ is equal to either $\nu_*$ or
$\nu_*/2$ \citep{lai08}. This model presents a promising possibility. However, the
problem is that we do not see a bending wave with the predicted frequency at the OVR.

In our simulations, a warp is usually located at the vertical resonance (or between
OVR and CR). However, the frequency of the warp either equals the frequency of the
star (if $r_m\approx r_{cr}$, see Sec. \ref{sec:fast warp}), has a low frequency (if
the simulation region is small, see Sec. \ref{sec:slow-small-region}), or there is no
large warp (if $r_m<r_{cr}$ and the simulation region is large, see Sec.
\ref{sec:slow-large-region}). From these possibilities we choose the most probable
one, where $r_m\approx r_{cr}$ and the high-amplitude warp corotates with the star.

We suggest that the upper QPO is connected with one of the high-frequency waves
described in Sec. \ref{sec:AMP-high frequency} (or with the rotation of hot spots
associated with this wave). The light from this high-frequency QPO can be reprocessed
by a warp, corotating with the star, so that the expected frequency of the lower QPO
is $\nu_l=\nu_u-\nu_*$. Such a model can explain the cases where the frequency
difference equals the frequency of the star. This approach is reminiscent of the
beat-frequency model (e.g., Lamb 1995), but in our model, we outline a possible
mechanism for the ``beat" frequency.


\subsubsection{Low-frequency, free bending oscillations in the disc}

Our simulations show that the tilted magnetosphere excites different types of waves in
the disc, including bending oscillations of the whole disc. These oscillations have
very low frequency, which can be of the order of the Keplerian frequency at the outer
radius of the disc, or even lower.

If the simulation region is relatively small, then these low-frequency waves couple
with the higher-frequency bending waves excited by the rotating magnetosphere, and a
slowly-rotating warp forms with the maximum of the amplitude at the OVR, and with the
frequency lower than the Keplerian frequency at the OVR.


One of the observed low-frequency QPOs has a frequency of $\nu_{LF}$, about 15 times
lower than that of the high-frequency QPOs (e.g., \citealt{vanderKlis06}). We can
estimate the size of the disc in cases where the frequency is equal to Keplerian
frequency at the edge of the disc. If the upper-frequency QPO is generated at distance
$r_u$ from the star (it can be, for example, the truncation radius of the disc), then
the Keplerian frequency at distance $r$ is $\nu(r)=\nu_u (r_u/r)^{3/2}$. For the
frequency ratio $\nu_u/\nu_{LF}=15$, we obtain the radius corresponding to $\nu_{LF}$,
$r_{LF}\approx 6.1 r_u$. Free oscillations of the inner part of the disc with this
radius are possible if the properties of the disc inside this radius differ from the
rest of the disc. We suggest that the magnetic-to-matter ratio may be larger inside
this region than at larger distances. For example, if accretion is provided by the
magneto-rotational instability (e.g., \citealt{balbus91}), then the turbulent cells in
the inner disc become strongly stretched in the azimuthal direction,  and the
azimuthal component of the field is generated due to rapid azimuthal rotation (e.g.,
\citealt{hawley00,armitage02,roma12}). This is only an example, and there may be other
reasons for different properties in the inner disc. There are, of course, other
possibilities. For example, \citet{lai99} suggested that the low-frequency
oscillations may be connected with a warp precessing around a magnetized star, which
is expected in cases where the rotational axes of the star and the disc are misaligned
(see also \citealt{pfeiffer04}).

To explain the even lower-frequency oscillations, such as $\nu_{\rm LL}\sim 1$Hz, we
need to consider an even larger disc. For example, if the high frequency
($\nu_u\approx 10^3$Hz) oscillations are generated at distance $r_u$, then the radius
corresponding to 1Hz Keplerian frequency is located at distance  $r_{\rm LL}\sim
(\nu_u/\nu_{\rm LL})^{2/3} r_u\sim 100 r_u$, which may be comparable to the overall
size of the disc.

\subsection{Application to Dwarf Nova oscillations}

The Dwarf Novae (DN) type of cataclysmic variables reveal two main types of
oscillations: the ``Dwarf Nova Oscillations" (DNOs) and the ``Quasi-Periodic
Oscillations" (QPOs) (e.g., \citealt{patterson81,warner04,pretorius06}). DNO
oscillations are observed exclusively during outbursts. They have relatively short
periods, ranging from about 7 to 40 s, and have a very high quality factor. They
display the period-luminosity correlation.

QPOs usually occur in dwarf novae outbursts, but are sometimes seen during the
quiescence. They have longer periods than the DNO, ranging from about 30 to 1000 s.
The QPO frequency is usually 15 times lower than the DNO frequency. QPOs are
characterized by broad humps in the power spectrum, indicating their short coherence
time of only a few cycles. \citet{robinson79} suggested that these oscillations are
produced in the accretion disc, possibly in a ``travelling spiral pattern" or in
``oscillating rings". Alternatively, they can be connected with the precession of a
warp \citep{lai99,pfeiffer04}.

In our model, we suggest that dwarf novae have a small magnetosphere, and the DNO
oscillations can be explained by trapped density waves, which develop in the inner
disc. They have a high frequency, which varies with the accretion rate (luminosity).
These oscillations are expected to be highly coherent. There are a few possibilities
for explaining QPOs (see Sec. \ref{sec:application-MP}). However, it is difficult to
explain the ratio $\nu_{DNO}/\nu_{QPO}\approx 15$, which is the same for various
accreting compact stars and also for black holes (e.g., \citealt{bel02,warner03}).

\section{Conclusions}
\label{sec:conclusions}

We performed the first global three-dimensional simulations of
waves in a disc excited by a rotating magnetized star with a tilted dipole magnetic
field. The main conclusions are the following:

\noindent\textbf{I. In cases where the inner disc approximately corotates with the
star ($r_m\approx r_{cr}$):}

 1. A strong one-armed bending wave (warp) forms near the corotation radius and
has its maximum amplitude at the radius of the vertical resonance.
   It corotates with the
magnetosphere and propagates out to the distance of $r\approx (2-2.5)r_m$ from the
star.

2. In the warp, the height of the center of mass of the disc,  $z_w$, can reach
$(5-10)\%$ of the distance to the star. This is in agreement with values found by
 \citet{terquem00} in their study for thin discs. However, the height of the warp $h_w$
(in our discs of finite thickness) can be as high as $30\%$.

3. Warps are excited by stars with dipole fields tilted at different angles,
$0^\circ<\theta<90^\circ$. However, the amplitude of the warp is larger in cases of
intermediate tilts of the dipole, $15^\circ < \theta < 60^\circ$.

4. Density waves are excited at different radii from the star, but their amplitudes
decrease beyond the distance of $r\approx 4$, which may be connected with the
relatively large thickness of the disc ($h/r\approx 0.1$). Bending waves propagate in
the whole simulation region, and have large amplitudes at different distances from the
star.

\smallskip

\noindent\textbf{II. In cases where the magnetosphere rotates more slowly than the
inner disc ($r_m <r_{cr}$):}

5. We observe  that the strong inner warp is absent. Only a much smaller warp
corotates with the star.
   Instead, a tightly wound rippled bending wave forms at the
disc-magnetosphere boundary and propagates out to the corotation radius.
   Beyond the
corotation radius, the wave is more ordered and less tightly wrapped, and it
propagates to large distances.

6. High-frequency \textit{trapped density waves} form in the inner disc, within the
maximum in the angular velocity distribution. These waves were predicted theoretically
by \citet{love07} and \citet{love09}.

7. High-frequency waves of another type -- \textit{inner bending waves} -- originate
at the disc-magnetosphere boundary during periods of unstable accretion.

8. Low-frequency global bending oscillations of the whole disc are excited in all
cases.  In the case of a smaller simulation region, we observe a slowly-rotating warp
with a maximum amplitude at the vertical resonance.  The low, sub-Keplerian frequency
results from the coupling between the frequency of the star and the global oscillation
mode of the disc. In the case of a relatively large simulation region, a
slowly-rotating warp does not form. Instead, global bending oscillations are observed.

\smallskip

\noindent\textbf{III. Some Applications:}

9. The formation of a \textit{high-amplitude warp} corotating with the star is
important for understanding obscuration of light in young stars (e.g.,
\citealt{bouvier03}). The formation of warps is expected in other types of stars. The
details of obscuration depend on the transparency of matter in the warp to the star's
radiation.

10. \textit{Trapped density waves and inner bending waves} can be responsible for
high-frequency QPOs in accreting millisecond pulsars, CVs, and for drifting ``period"
in CTTSs.  Trapped density waves appear in a wide variety of situations, where the
magnetosphere rotates slower than the inner disc.

11. Inner high-frequency waves can determine the position of funnel streams and
\textit{moving spots} on the surface of the star. Hence, they can leave an imprint on
QPO frequencies associated with moving spots discussed e.g. by \citet{bac10}.

12. Global disc oscillations and slowly-rotating warps can be responsible for
low-frequency QPOs in different stars.

\section*{Acknowledgments}

The authors thank A. Blinova for help preparing this paper. Resources supporting this
work were provided by the NASA High-End Computing (HEC) Program through the NASA
Advanced Supercomputing (NAS) Division at Ames Research Center and the NASA Center for
Computational Sciences (NCCS) at Goddard Space Flight Center.  The research was
supported by NASA grants NNX10AF63G, NNX11AF33G and NSF grant AST-1008636. AVK and GVU
were supported in part by RFBR grants 12-02-00687 and 12-01-0060.

\begin{appendix}

\section{Waves in a Thin Keplerian Disc} \label{sec:Appendix-allwaves-modes}

Here, we briefly summarize the key aspects of the low mode-number waves in thin
Keplerian discs.

\begin{figure}
\centering
\includegraphics[width=8cm]{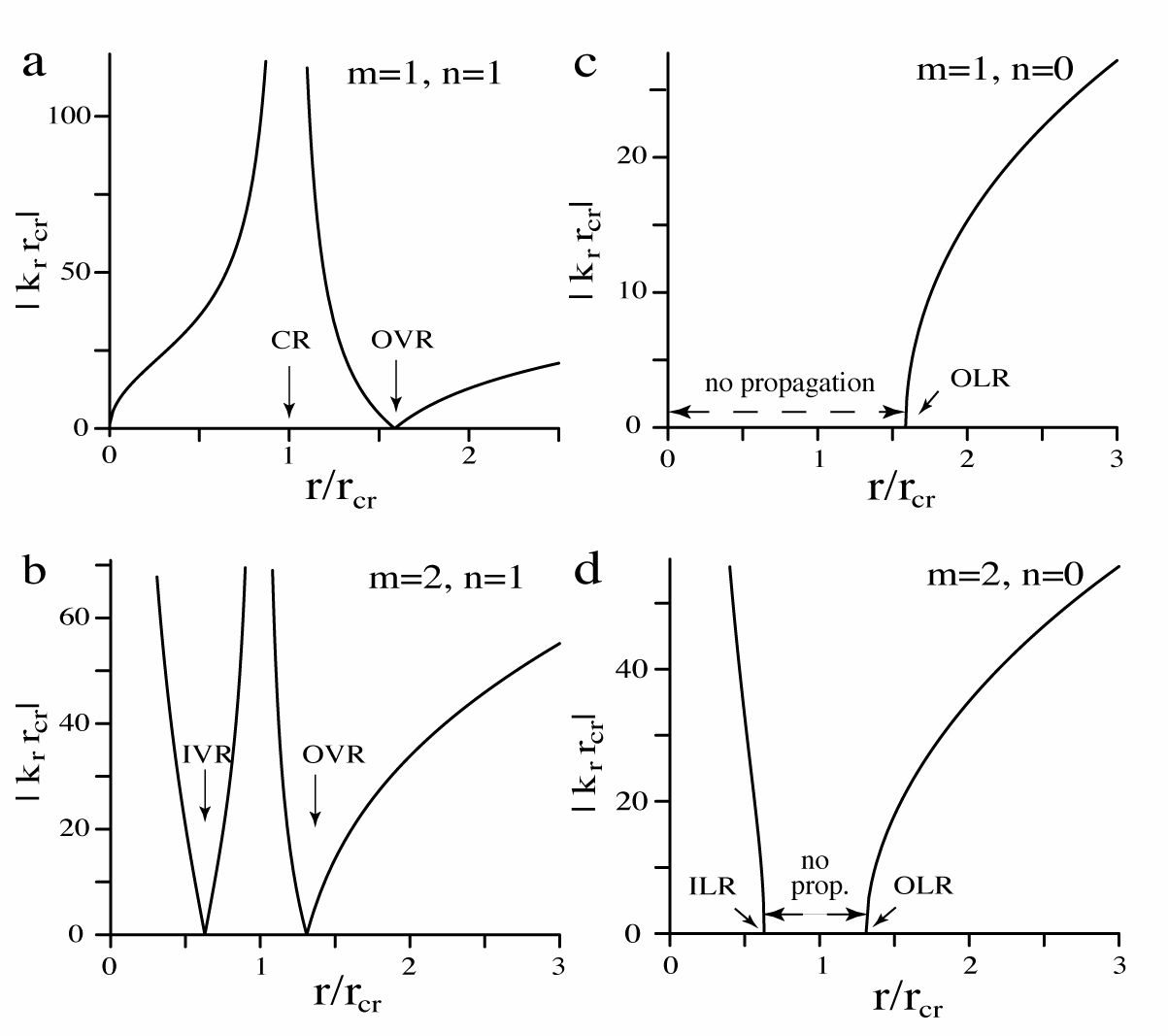}
\caption{Radial wavenumber $k_r$ as a function  of radius $r$ for a thin Keplerian
disc with $h/r =c_s/(r\Omega) =0.05$, where $r_{cr}$ is the corotation radius.  Panels
(a) and (b)  are for the bending waves, while (c) and (d) are for the density waves. }
\end{figure}

\begin{figure*}
\centering
\includegraphics[width=4.0cm]{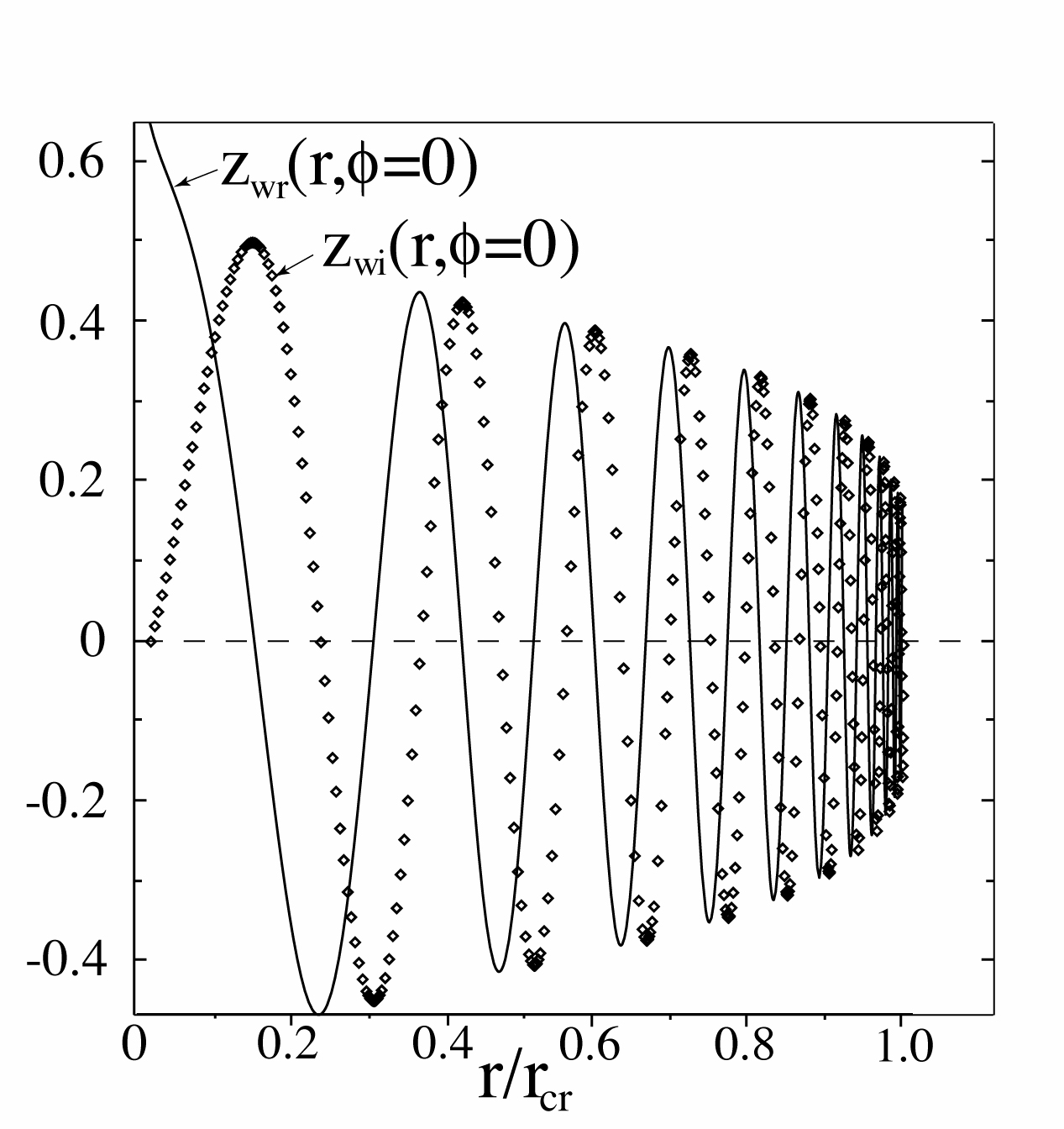}
\includegraphics[width=4.0cm]{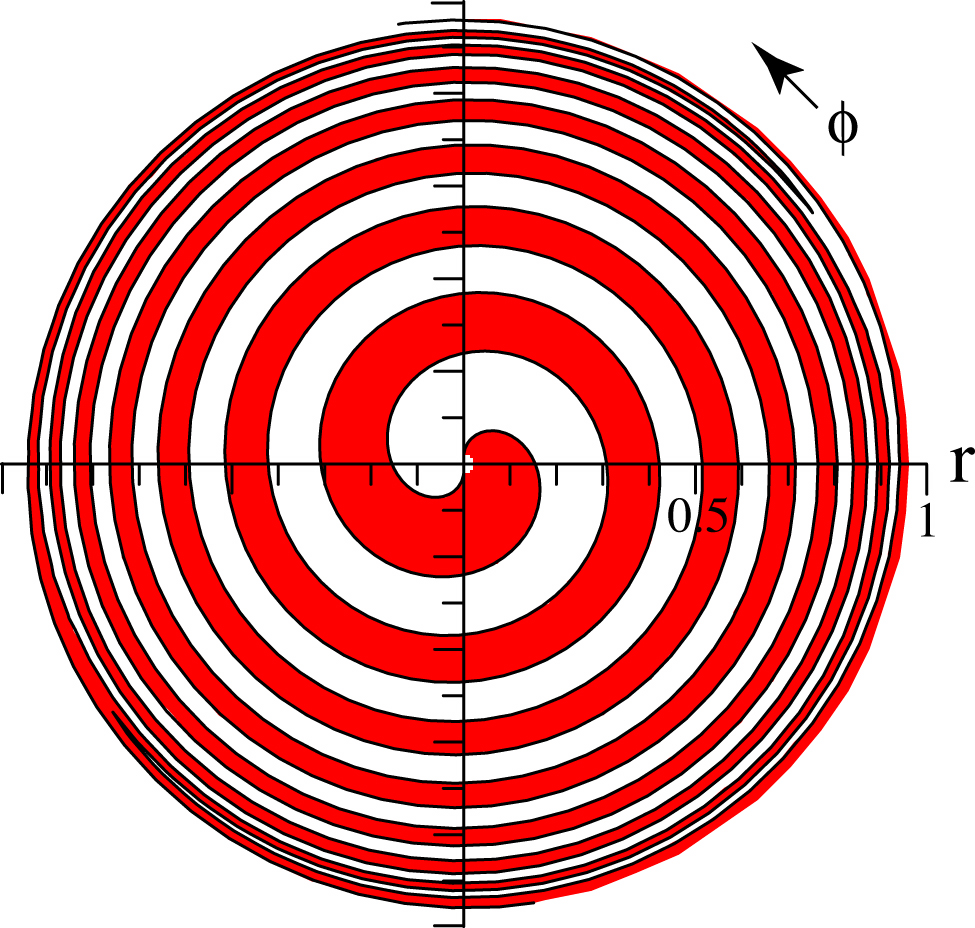}
\includegraphics[width=4.0cm]{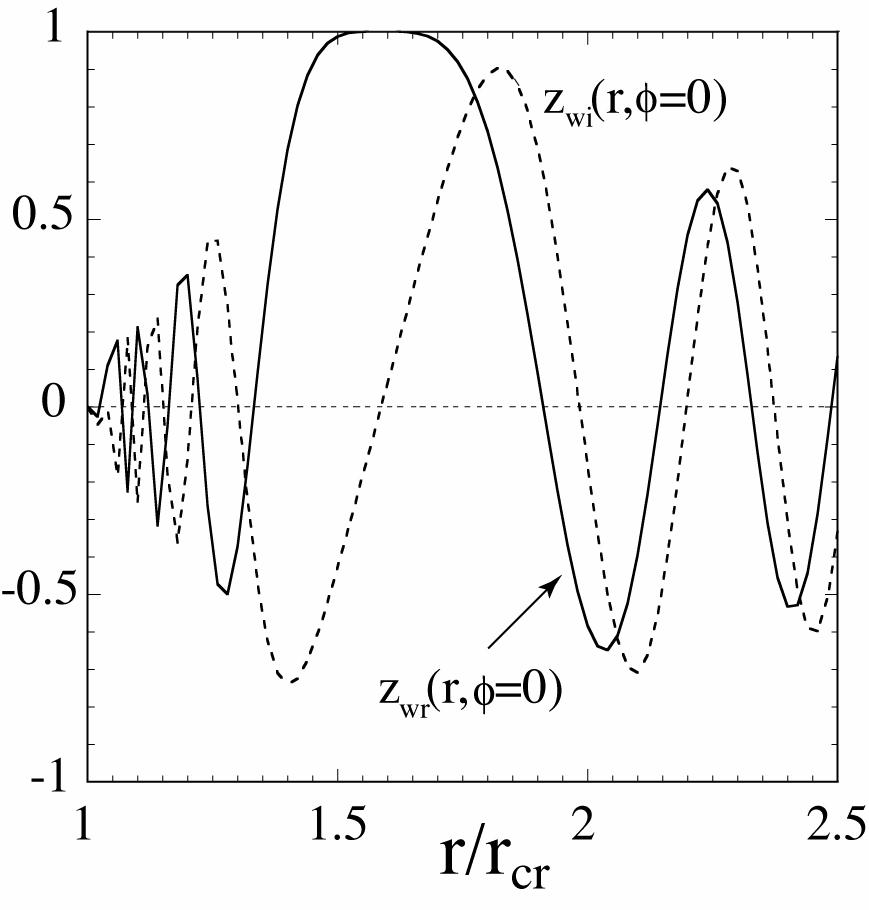}
\includegraphics[width=4.0cm]{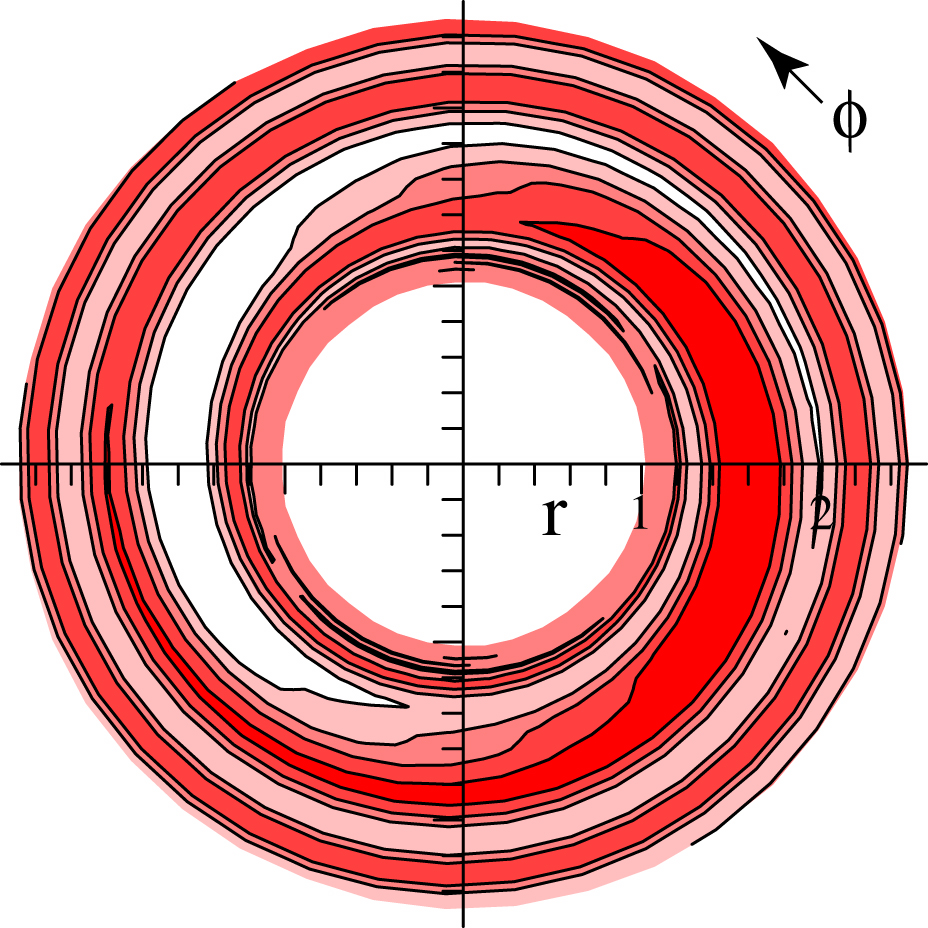}
\caption{From left to right: (1) Profiles of the real and imaginary parts of the
displacement of the disc $z_w(r,\phi=0)$ for the `inner warp', $r/r_{cr} < 1$,
assuming $c_s/(r\Omega)=0.05$. (2) Vertical displacement of the disc $z_w(r,\phi)$ for
the `inner warp', $r/r_{cr} < 1$, assuming $c_s/(r\Omega)=0.05$. Here, white =
negative and red = positive. (3) Real and imaginary parts of the disc displacement
$z_w$ for the `outer warp' from equation A1. (4) Vertical displacement of the disc
$z_w(r,\phi)$ for the `outer warp' for $1<r/r_{cr}<2.5$ assuming
$c_s/(r\Omega)=0.05$.} \label{Appfig:allwarps}
\end{figure*}

\subsection{m=1 Bending Wave} \label{sec:Appendix-warping-modes}

   For a one-armed ($m=1$) bending wave
($n=1$)  of a Keplerian disc ($\kappa=\Omega$ and $\Omega_\perp =\Omega$) equation (4)
can be written as
\begin{equation}
\left\{ \left({k_r c_s\over \Omega}\right)^2-\left[\left({\tilde\omega\over
\Omega}\right)^2 -1\right]\left[1-\left({\Omega\over \tilde\omega}\right)^2\right]
\right\}z_w=0,
\end{equation}
where $\tilde\omega =\omega-\Omega$ and the other parameters are defined in Sec. 2.

We find that the WKBJ equation (3) is a good approximation to the full wave solution
{\it except} for the region where $|k_r r_{cr}|$ is small or the wavelength is long.
Specifically, the approximation is reasonable for $|k_r r_{cr} |\gtrsim 4.94$.
   For $|k_r r_{cr} |<5$, equation (A1) is solved as a differential
equation for $z_w$ obtained by setting $k_r^2 \rightarrow -d^2/dr^2$.

   Panel (a) of  Figure A1 shows the radial  dependence of $|k_r r_{cr}|$ for a disc
with constant fractional half-thickness $h/r = c_s/(r\Omega) = 0.05$ and $\omega =
\Omega(r_{cr})$.
    The corotation resonance where $\tilde\omega=0$ and $k_r^2
\rightarrow \infty$ is at $r=r_{cr}$.
    The vertical/Lindblad resonance
where $\tilde\omega =\Omega$ and $k_r^2 =0$ is at $r=2^{2/3}r_{cr} \approx 1.59
r_{cr}$.
    For $r/r_{cr} \rightarrow 0,$ $k_r r_{cr} \rightarrow 2 (r/h)\sqrt{r/r_{cr}}$.
For $r/r_{cr} \rightarrow 1,$ $k_r r_{cr} \rightarrow (r/h)|(r/r_{cr})^{3/2}-1|^{-1}$.
  For $r/r_{cr} \rightarrow \infty,$ $k_r r_{cr} \rightarrow (r/h)\sqrt{r/r_{cr}}$.

We refer to the warp in the range  $r/r_{cr} <1$ as the `inner warp.' For most of the
range $r/r_{cr} <1$ (excluding very small $r/r_{cr}$), the amplitude of the inner warp
is well approximated by equation (A2). Fig. A2 (left panel) shows the real and
imaginary parts of the warp profile at $\phi=0$. Next panel to the right shows the top
view of $\Re[z_w(r,\phi)]$ of the inner warp.
      The warp becomes very tightly wrapped
as $r/r_{cr}$ increases towards unity.
       Approximately,
$\phi \sim - (2r/3h) \ln[r_{cr}/(r_{cr}-r)]$ and $|z_w| \sim \sqrt{r_{cr}-r}$ as $r
\rightarrow r_{cr}$.

   In the range $r/r_{cr}>0$ we refer to warp as the `outer warp'.
Fig. A2 (3rd panel) shows the real and imaginary parts of $z_w$ at $\phi=0$. As $r
\rightarrow r_{cr}$.$|z_w| \sim \sqrt{r-r_{cr}}$ while for $r\gg r_{cr}$ $|z_w| \sim
r^{-1/4}$.
 Fig. A2 (right panel) shows the top view of $\Re [z_w(r,\phi)]$.

\subsection{ m=2 Bending Wave}

   Using the WKBJ approximation to
equation (2) for a two-armed ($m=2$) warp ($n=1$) of a Keplerian disc gives
\begin{equation}
|k_r r_{cr} |=  {r\over h}{1\over \bar r}\bigg[[(\bar r^{3/2}-1)^2-{1/4}][4-(\bar
r^{3/2}-1)^{-2}]\bigg]~,
\end{equation}
where $\bar r \equiv r/r_{cr}$ with $r_{cr}$ the corotation radius, so that
$\omega=m\Omega(r_{cr}) $.
  Note that $k_r=0$ at the inner and outer vertical
resonances (IVR and OVR) at $\bar r =(1\pm 1/2)^{2/3} \approx 0.630,~ 1.31$, and that
$|k_r |\rightarrow \infty$ at the corotation resonance at   $\bar r = 1$. Panel (b) of
Fig. A1 shows the radial dependence of $k_r$  for a sample case.

\subsection{In-plane modes, $n=0$ (Density Waves)}

  The WKBJ solution of equation (2) for $n=0$, $m=1,~2$, and
a Keplerian disc gives
  \begin{equation}
  \tilde\omega^2=\Omega^2 +k_r^2c_s^2~,
  \end{equation}
 where we have set $\kappa=\Omega$.
    For a disc with $c_s/(r\Omega)=h/r =$ const,
one finds
\begin{equation}
|k_r r_{cr}| =  (r/h)(1/\bar r)\left[m^2(\bar r^{3/2}-1)^2 -1\right]^{1/2}~,
\end{equation}
where $\bar r \equiv r/r_{cr}$ with $r_{cr}$ being the corotation radius,
$\omega=m\Omega(r_{cr})$.
    For the  one-armed mode ($m=1$), there are real values
of $k_r$ only for $r/r_{cr} \geq 2^{2/3}\approx 1.59$, which is the outer Lindblad
resonance.

    For the two-armed mode ($m=2$), there are real
values of $k_r$ only for $r/r_{cr}\leq 2^{-2/3}\approx 0.63$ (the inner Lindblad
resonance) and for $r/r_{cr}  \geq 1.5^{2/3} \approx 1.311$ (the outer Lindblad
resonance).
    Panels (c) and (d) show the
the radial dependencies of $k_r$ for the two cases.

\section{Radially trapped modes}
\label{sec:Appendix-trapped modes}

A radially localized, in-plane instability of discs for $m=1,~2,..$ know as the
`Rossby wave instability' (RWI) may occur near regions where there is an extremum (as
a function of $r$) of the entropy and/or the potential vorticity $\hat{\bf z}\cdot
(\nabla \times {\bf v})/\Sigma$, where ${\bf v}$ is the disc velocity and $\Sigma$ is
the disc's surface mass density
\citep{lovelace-Li99,li-finn-lovelace00,love09,love07}; see also \citet{lovelace78}.
Radially trapped modes in discs around black holes were analyzed earlier by
\citet{nowak91}.
    Of particular interest here is the RWI in the non-Keplerian region of a disc (\citealt{love09}; hereafter LTR09)
    which occurs when a disc encounter the magnetosphere of a  slowly-rotating star.
    An example of such a rotation curve is shown in Figure B1.
     For given profiles of $v_\phi$, $\rho$, etc. and
a given $m$,   the value of the growth rate is found by solving for the `ground state'
solution of Schr\"odinger-like equation for the enthalpy perturbation $\psi$,
\begin{equation}
{d^2 \psi \over dr^2}= U(r | \omega)\psi~,
\end{equation}
where $U(r|\omega)$ is an effective potential (LTR09) shown for a sample case in
Figure B2.
  The ground state is found by varying both the real
and imaginary parts of the mode frequency $\omega$ so as to give the $\psi(r)$ most
deeply trapped in the potential well.  For typical profiles and $m=1$, the growth rate
is $\omega_i\sim 0.1 (v_\phi/r)_R$ and the real part of the frequency is
$\omega_r\approx (v_\phi/r)_R$, where the $R-$subscript indicates evaluation at the
resonant radius $r_R$.
    The radial width of the mode is $\Delta r/r_R \sim 0.05$.

     The linear growth of the Rossby wave is predicted
to saturate when the azimuthal frequency of trapping of a fluid particle in the trough
of the wave $\omega_T$ grows to a value equal to the growth rate $\omega_i$ as
discussed by LTR09.
  For the $m=1$ mode the saturation level is
estimated as $|\delta \rho|/\rho \sim 1.4(\omega_i/\omega_r)^2(r/h)$ (LTR09).

\begin{figure}
\centering
\includegraphics[width=4cm]{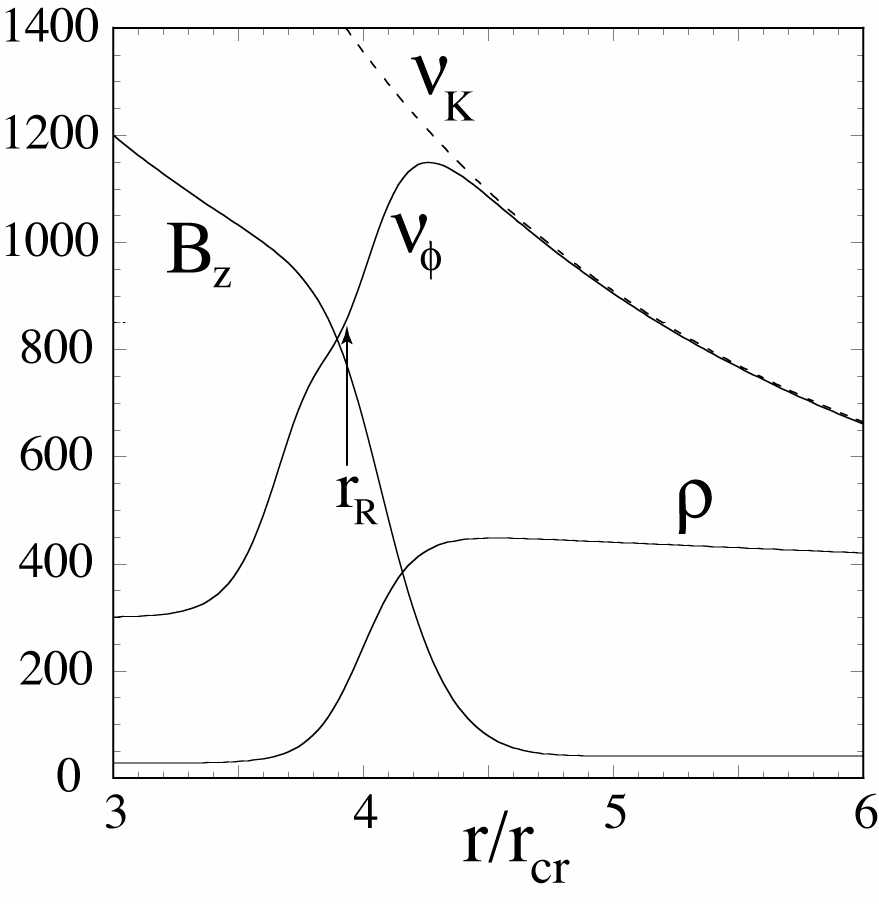}
\includegraphics[width=4cm]{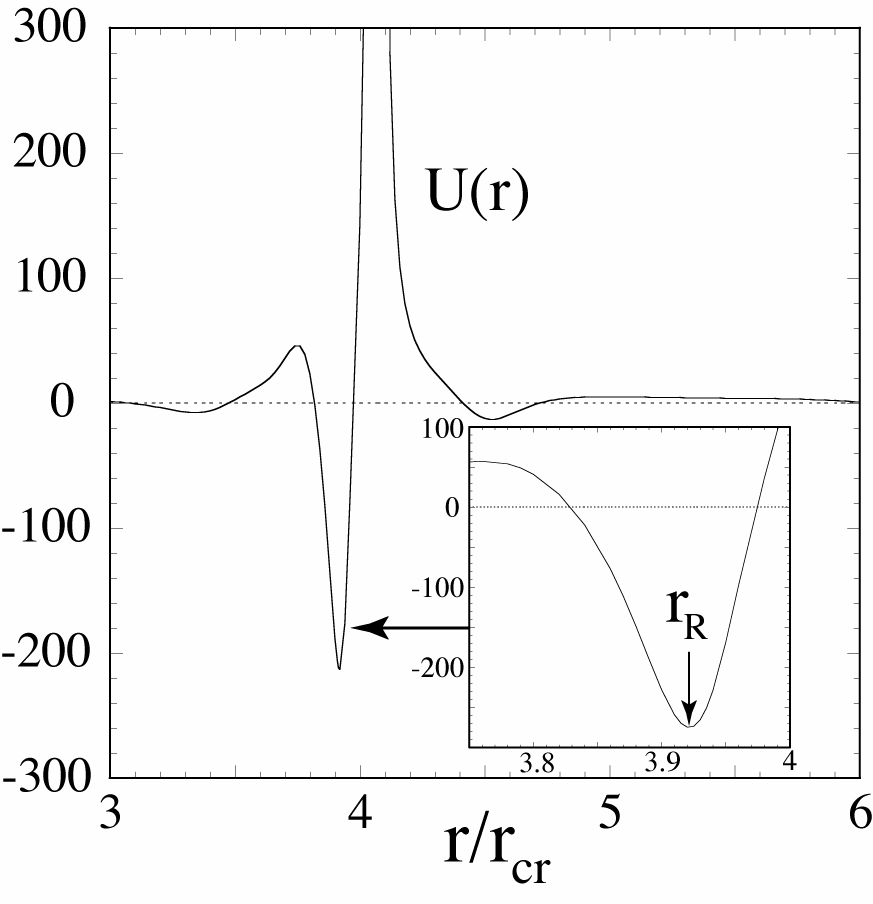}
\caption{\textit{Left panel}: Midplane radial dependencies of the azimuthal disc
velocity $v_\phi$, magnetic field $B_z$, and density $\rho$, where $v_K$ is the
Keplerian velocity.   The vertical arrow indicating $r_R$ is the location of the
bottom of a potential well $U(r)$ shown in right panel (from LTR09). \textit{Right
panel}: Effective potential $U(r)$ for the profiles shown in Figure D1 from LTR09.}
\label{Appfig:trapped-all}
\end{figure}

\end{appendix}

\end{document}